\documentstyle[preprint,aps,eqsecnum,floats,epsfig]{revtex}

\newcommand{\beq}{\begin{equation}}
\newcommand{\eeq}{\end{equation}}
\newcommand{\beqa}{\begin{eqnarray}}
\newcommand{\eeqa}{\end{eqnarray}}

\newcommand{\bxnn}{\mbox{$b \to X\,\nu\,\bar\nu$}}
\newcommand{\bqnn}{\mbox{$b \to X_q\,\nu\,\bar\nu$}}

\newcommand{\bctn}{\mbox{$b \to X_c\,\tau\,\bar\nu$}}

\newcommand{\Btn}{\mbox{$B^- \to \tau\,\bar\nu$}}

\newcommand{\Bln}{\mbox{$B^- \to \,l\, \bar\nu $}}

\newcommand{\Btt}{\mbox{$B \to \,\tau^+\,\tau^-$}}
\newcommand{\Bqtt}{\mbox{$B_q \to \,\tau^+\,\tau^-$}}
\newcommand{\Bdtt}{\mbox{$B_d \to \,\tau^+\,\tau^-$}}
\newcommand{\Bstt}{\mbox{$B_s \to \,\tau^+\,\tau^-$}}

\newcommand{\Bqll}{\mbox{$B_q \to  \,l^+\,l^-$}}

\newcommand{\Bqmm}{\mbox{$B_q \to \,\mu^+\,\mu^-$}}

\newcommand{\bstt}{\mbox{$b \to X_s\,\tau^+\,\tau^-$}}

\newcommand{\bxtt}{\mbox{$b \to X\,\tau^+\,\tau^-$}}
\newcommand{\bxll}{\mbox{$b \to X\, l^+\, l^-$}}

\newcommand{\bsll}{\mbox{$b \to X_s\,l^+\,l^-$}}

\newcommand{\bsmm}{\mbox{$b \to X_s\,\mu^+\,\mu^-$}}

\newcommand{\BR}{\mbox{${\rm BR}$}}

\newcommand{\la}{\mbox{$\lambda$}}
\newcommand{\lp}{\mbox{$\lambda^\prime$}}

\newcommand{\gsim}{\mbox{${~\raise.15em\hbox{$>$}\kern-.85em
          \lower.35em\hbox{$\sim$}~}$}}
\newcommand{\lsim}{\mbox{${~\raise.15em\hbox{$<$}\kern-.85em
          \lower.35em\hbox{$\sim$}~}$}}
\newcommand{\Lbs}{\mbox{${\scriptstyle \not \!\!\; L}$}}
\newcommand{\Bbs}{\mbox{${\scriptstyle \not \!\!\; B}$}}

\def\Re{{\rm Re}}
\def\Im{{\rm Im}}

\def\mh{\hat{m}}

\def\mlh{\hat{m}_l}

\def\mch{\hat{m}_c}
\def\msh{\hat{m}_s}

\def\mvh{\hat{M}_V}
\def\sh{\hat{s}}

\def\a{\alpha}

\def\g{\gamma}
\def\G{\Gamma}

\def\l{\lambda}

\def\m{\mu}
\def\n{\nu}
\def\p{\pi}

\def\t{\tau}

\def\nnu{\nonumber}

\def\fba{A_{\rm {FB}}(\sh)}

\def\mh{\hat{m}}
\def\mlh{\hat{m}_l}

\def\mch{\hat{m}_c}
\def\msh{\hat{m}_s}

\def\sh{\hat{s}}

\def\a{\alpha}

\def\ea{{\it et al.\/}}

\def\npb#1{Nucl.\ Phys.\ {\bf B #1}}
\def\plb#1{Phys.\ Lett.\ {\bf B #1}}
\def\prd#1{Phys.\ Rev.\ {\bf D #1}}
\def\prl#1{Phys.\ Rev.\ Lett. {\bf #1}}

\def\zpc#1{Z.~Phys.\ {\bf C #1}}

\def\m{\mu}
\def\GeV{{\rm{GeV}}}
\def\MeV{{\rm{MeV}}}

\def\Vec#1{\mbox{\bf{#1}}}
\def\eN{\Vec{e}_{\rm{N}}}
\def\eL{\Vec{e}_{\rm{L}}}
\def\eT{\Vec{e}_{\rm{T}}}

\def\lop{{\cal P}_{\rm L}}
\def\trp{{\cal P}_{\rm T}}
\def\nop{{\cal P}_{\rm N}}

\def\c9eff*{c_9^{\rm{eff}*}}
\begin{document}

\draft

{\tighten
\preprint{\vbox{\hbox{WIS-97/24/Jul-PH}
                \hbox{hep-ph/9707371} }}

\title{Searching for New Physics in Rare  $B \to \tau$ Decays}

\author{Dafne Guetta\,$^{a,1}$ and   Enrico Nardi\,$^{b,2}$ }

\footnotetext{
\begin{tabular}{ll}
E-mail addresses: & $^1$guetta@bo.infn.it \\
& $^2$nardie@vxcern.cern.ch
\end{tabular} }

\address{ \vbox{\vskip 0.truecm}
  $^a$Dipartimento di Fisica \\
Universit\'a degli Studi di Bologna \\
Via Irnerio 46, I-{\it 40126}  \  Bologna, \ Italy \\
\vbox{\vskip 0.truecm}
  $^b$ Departamento de F\'\i sica \\
Universidad de Antioquia \\
A.A. {\it 1226} \ Medell\'\i n,  \ Colombia }

\maketitle

\begin{abstract}%
The rare decays \mbox{$B^- \to \tau\,\bar\nu$},
\mbox{$B \to \,\tau^+\,\tau^-$},
\mbox{$b \to X\,\nu\,\bar\nu$}\ and
\mbox{$b \to X\,\,\tau^+\,\tau^-$}\
all contain third generation leptons in the final state, and hence are
sensitive to new physics that couples more strongly to the third
family. We present model independent expressions for these decays that
can be useful to study several types of new physics effects. We
concentrate on supersymmetric models without R-parity and without
lepton number. We also assume a horizontal $U(1)$ symmetry with
fermion horizontal charges chosen to explain the magnitude of fermion
masses and quark mixing angles. This allows us to estimate the order
of magnitude of the new effects, and to derive numerical predictions
for the various decay rates and for the forward-backward asymmetry
and the $\tau$ polarization components measurable in  \mbox{$b \to
X\,\,\tau^+\,\tau^-$}. In some cases the branching ratios are
enhanced by more than one order of magnitude, rendering
foreseeable their detection at upcoming $B$-factories.
We also discuss how a measurement of  asymmetries in
\mbox{$b \to X\,\,\tau^+\,\tau^-$}\ can be crucial in distinguishing
between different sources of new physics. 
\end{abstract}
} % end tighten

\bigskip
\leftline{ PACS number(s):  13.20.-v, 13.20.He, 14.60.Fg, 13.88.+e, 12.60.Jv}

\newpage

 {\tighten    %%%%%%    For 35 or 27 pages preprints

\section{Introduction}

The standard model (SM) of the strong and electroweak interactions provides a
successful description of all the phenomena involving the known elementary
particles.  However, experimental results involving fermions of the third
generation are far less precise than for the first two generations.
{}From the theoretical point of view, several models of new
physics predict larger deviations from the SM 
for processes involving third generation fermions \cite{Topcolor}.
This is also the case in a class of supersymmetric (SUSY)
models without R-parity \cite{susynoR}  where violation of
lepton ($L$) and baryon ($B$) number couples more strongly
to the heavier fermions \cite{BGNN1,YossiYuval,Valery,BGNN2}.

In recent years, the experiments at the CERN $e^+ e^- $ collider LEP
have provided us with most of the results on $b$ decays into the
third generation leptons $\tau$ and $\nu_\tau\,$.
This is because the LEP environment has the advantage over
symmetric $B$-factories (like CLEO) or  hadron colliders (like CDF)
of allowing accurate measurements of the missing energy
associated with primary $b\to \nu_\tau$ or  secondary
$ b\to \tau \to \nu_\tau$  final  state neutrinos.
In this way, decay modes yielding a  missing energy spectrum harder
than the usual semileptonic decay can be effectively measured or
constrained. At LEP,  measurements of  an excess of events
over the semileptonic background  with missing energy between
10  and 30 GeV \cite{ALEPH,L3a,OPAL}
was interpreted as the signature of the decay
$B\to X_c\,\tau\,\bar\nu_\tau$ followed
by $\tau\to\nu\,X\,$. This yielded $ \BR(\bctn) = 2.68 \pm 0.34 \% \,$
in agreement with the SM prediction $\BR(\bctn)=2.30\pm0.25\%$
\cite{FLNN}.
Using a large missing energy tag the L3 Collaboration set the 90\%
confidence level upper limit on the exclusive leptonic decay \Btn \cite{L3b}
\beq \label{Btnlimit}
  \BR(\Btn) < 5.7 \times 10^{-4} \,.
\eeq
In \cite{bsnn}  it was discussed how similar analyses
can yield a limit on  the flavor changing decay $\bxnn\,$. Based of the full
LEP--I data sample, the ALEPH Collaboration derived a
preliminary 90\% confidence level limit on this decay mode \cite{warsaw}
\beq \label{bsnnlimit}
  \BR(\bxnn) < 7.7 \times 10^{-4} \,.
\eeq
Being only one order of magnitude above the SM predictions,
the limits (\ref{Btnlimit}) and (\ref{bsnnlimit}) imply
strong constraints on several models of new physics
\cite{bsnn,bctn-mhdm}.

The tight limits on very large missing energy events
($E_{\rm miss}>35\,$GeV ) in $b$ decays reported by the ALEPH
collaboration \cite{ALEPH} allowed to estimate order of magnitude 
bounds on \BR(\Bdtt) and  \BR(\Bstt) \cite{btt}.
In Ref.~\cite{btt} it was also argued that a weak upper limit 
on the branching ratio for  \bxtt\ of the order of the semitauonic
branching ratio is implied by the LEP  
missing energy measurements. The limits were estimated as~\cite{btt}
\beqa \label{Bsdxttlimits}
 \BR(\Bdtt) &<& 1.5 \times 10^{-2} \nnu \\
 \BR(\Bstt) &<& 5.0 \times 10^{-2}  \nnu \\
 \BR(\bxtt) &<& 5.0 \times 10^{-2}\,.
\eeqa
Even if several orders of magnitude above the SM
rates, these figures still yield 
bounds on some new physics parameters  which 
are unconstrained by other processes \cite{btt}.

In the near future, experiments at $B$-factories will reach a much higher
sensitivity in the study of $b\to \tau\,, \nu_\tau$ decays.
A measurement of \Btn\  appears to be accessible even at the low SM rate.
With refined experimental techniques and after few years of run,
the experimentally very challenging decay \bxnn\
could also be measured, at least in some exclusive decay channel.
Because of the even lower rates and of the expected low efficiency in
$\tau$ identification, the decays \Bdtt\ and \bxtt\
might be out of the reach of experiments like BaBar and BELLE if their
rates are  at the SM level. However, as we will discuss,
some  new physics models predict  decay rates more than one order 
of magnitude above the SM. 
Therefore, dedicated studies of $b\to \tau\,,\nu_\tau$ 
decays at $B$-factories  represent 
a powerful tool for detecting signals of new physics.

The paper is organized as follows.
In section II we introduce general four fermion
amplitudes for the decays  $\Bln\,$, $\Bqll$  and
$\bqnn$ ($q=d,s$)  and we give the results for
the various branching ratios. Since our approach is essentially model
independent, it is  well suited to study  different types of
new physics contributions.

In section III we concentrate on the decay \bxll. To take into
account the effects of new physics, the standard basis of operators
contributing to the effective Hamiltonian of the decay \cite{Witten}
has to be enlarged. We generalize it by introducing a set of
operators for the right-handed flavor changing current
$\bar s_R\gamma_\mu b_R$ together with a set of new scalar operators.
These new effective operators  arise in several
new physics models, like SUSY models without R-parity \cite{susynoR},
models with leptoquarks \cite{LQ}, left-right
symmetric models \cite{LRrizzo}, etc.
We study the various observables measurable in the decay: the inclusive
rate, the forward-backward asymmetry and the $\tau$ polarization asymmetries.
Since for the decay channel $\bxtt\,$  
the effect of the $\tau$  mass is non negligible and  the average
energy of the final hadronic system is not very large,
in our computation we retain all the fermion masses.
We next apply our results to the
study of  SUSY models without R-parity and without $L$ number.
The theoretical framework is  presented in
section IV.  Since in these models the values of the various R-parity
violating couplings is not determined, without
further theoretical input no numerical prediction
of the corresponding effects on  $B$ decays is possible.
In order to estimate  the new physics effects,
we appeal to models where the magnitude of the fermion masses and
CKM mixing angles is explained by assuming some horizontal $U(1)$ symmetry.
This framework provides us with additional theoretical constraints
yielding a set of numerical predictions for the various $L$ violating
couplings, and allowing for order of magnitude estimates of
the various decay rates.

In section V we present a numerical analysis of two representative
models and we discuss the results.
In particular, we compare our estimates  for decays involving the
transition $b\to \tau$  with the corresponding  decays involving
muons, and we confront the predictions of our new physics models
with the SM. Finally, section VI contains the summary and our conclusions.
The set of input parameters used in the numerical
analysis is collected in an Appendix.

\section{The decays \Btn, \Btt\ and \bxnn.}

\bigskip

The most general non-derivative effective four-fermion interaction
involving a $b$ quark,
a $q=d\,,s$ or $u$ quark, and a pair of leptons $\ell$ and $\ell'$
can be written  in the form
\beq \label{heff}
{\cal H}^{qb}_{\rm eff} = - G_F\,
 \sum_a \, (\bar q\,\Gamma_a\,b)\,
\left(\bar \ell\,[C_a\, \Gamma_a\,  +
 {C_a}' \,\Gamma_a\,\gamma_5\,] \ell'\right)
\eeq
where
$\Gamma_a=\{I,\gamma_5,\gamma^\mu,\gamma^\mu\gamma_5,\sigma^{\mu\nu}\}$
with $a=\{S,P,V,A,T\}\,$  the standard basis of operators of
the Clifford algebra. In (\ref{heff}) we  have factored out the
Fermi constant $G_F$ so that all the coefficients $C_a$ and
$C'_a$ are dimensionless. Even in the presence of new physics,
most of the rare $B$ decays depend only on a subset of the
operators in (\ref{heff}). This is due to the fact that for
purely leptonic $B$ decays, several matrix elements of the quark
operators vanish. Assuming that neutrinos are described by two
component left-handed spinor fields,  the number of relevant
operators is further reduced when neutrinos appear in the final
state.\footnote{We assume that even in the presence of new
physics, the neutrinos produced in $B$ decays are mainly the SM
ones. This is not a strong assumption, since light right-handed
neutrinos are theoretically disfavored. If the `SM
neutrinos' have  non-vanishing masses, operators that vanish in
the massless limit are suppressed at least as $m_\nu/m_b$ and
hence always negligible.} We will now list the general
expressions for the different decays.

\subsection{The decay \Bln. }

The decay \Bln\ is described by the effective Hamiltonian
(\ref{heff}) with $q=u\,$, $\ell=l_i\,$ and $\ell'=\nu_j $ where
$i,j=1,2,3$ correspond to the different lepton flavors.
In the presence of new physics
(for example in SUSY models without R-parity)
$i\neq j$ is an open possibility.  Since
final state neutrinos are not detected, in these cases
a sum has to be taken over all the allowed decay modes.

The general amplitude for this decay involves a set of
matrix elements  $\langle0|\,\bar q\,\Gamma_a\,b\,|B\rangle\,$.
They  vanish for the parity-even operators
$\Gamma_S=I$ and $\Gamma_V=\gamma^\mu$ due to
the pseudoscalar nature of the $B$ meson.
The tensor operator $\Gamma_T=\sigma_{\mu\nu}$ is
antisymmetric in the Lorentz indices, and hence 
its matrix element must vanish as well, since the only
available four-vector is the momentum  $p_B^\mu\,$ of the
$B$ meson. Therefore, only the matrix elements of the
pseudoscalar and axial-vector operators
contribute. They  are given by
the PCAC (partial conserved axial current) relations
\beqa \label{PCAC}
\langle0|\,\bar u\,\gamma^\mu\gamma_5\,b\,|B^-\rangle
&=& i f_B\, p^\mu_B \,,
  \nonumber \\*
\langle0|\, \bar u\,\gamma_5\,b\, |B^-\rangle &=&
- i f_B\, {m_B^2\over m_b+m_u} \simeq  - i f_B\,\frac{m^2_B}{m_b} \,.
\eeqa
Under the assumption of two-component left-handed
neutrinos $\nu_L = P_L\nu\,$,
with $P_{L} = \frac{1}{2} (1 - \gamma_5)\,$,
we further have $\bar l \g_5 \nu_L = - \bar l \nu_L$
and $\bar l \g_\m \g_5 \nu_L = - \bar l \g_\mu \nu_L\,$.
For on-shell final state leptons, the latter operator contracted with
the $B$ meson four-momentum
$p_B^\mu=k_l^\mu+k_{\bar\nu}^\mu$ yields
$p^\m_B\, ( \bar l \g_\mu \nu_L\,) = m_l\, ( \bar l\nu_L\,)\,$.
Hence the  amplitude for the  \Bln\ decay reads
\beq \label{BlnAmpl}
{\cal{A}}^{l\bar\nu} =  i\, f_{B}\, m_{B}\, G_{F} \left[
({C_A^{l\bar\nu}}-{C_A^{l\bar\nu}}')\frac{m_l}{m_B} -
(C_P^{l\bar\nu}-{C_P^{l\bar\nu}}')
\right]\left( \bar l\,\nu_{L} \right)\,.
\eeq
The corresponding expression for the decay rate is
\beq \label{BlnRate}
\BR(B^- \to l\bar\nu) = f^2_B \tau_B\,
\frac{G^2_F m_B m_l^2}{16\pi}\left[1 - \frac{m^2_l}{m^2_B}\right]^2
\sum_i\,\left|(C_A^{l\bar\nu_i}-{C_A^{l\bar\nu_i}}') -
\frac{m_B}{m_l}
({C_P^{l\bar\nu_i}}-{C_P^{l\bar\nu_i}}')\right|^2 \,,
\eeq
where the sum over the index $i$
accounts for possible $\nu_i \neq \nu_l$  decay channels.
For $l=\tau\,$ new physics can induce sizeable  enhancements 
over the SM rate (this can occur for example in multi Higgs doublet 
models where new contributions
arise from charged Higgs exchange diagrams  \cite{Hou}). However, 
a large  theoretical uncertainty in predicting the branching
ratios is associated with the present poor knowledge of $f_B\,$.
Therefore, it could be difficult to identify unambiguously new 
physics effects in this decay.

In the SM  $C_P^{l\bar\nu} = {C_P^{l\bar\nu}}' =0 $ and
\beq
\Big[{C_A^{l\bar\nu}}-{C_A^{l\bar\nu}}'\Big]_{\rm SM}
=-\sqrt{2}\,V_{ub}\,.
\eeq
Using the set of reference parameters listed in the Appendix,
for the SM branching ratio we find
$\BR^{\rm SM}(\Btn) = 7.1\times 10^{-5}\,$.

\subsection{The decay \Bqll. }

A detailed analysis of the decay \Bqll\
and of the possible types of new physics contribution
was presented in \cite{btt}. For on shell $\tau $'s,
$\>p^\m_B \> (\bar l \g_\mu l) = (k_{\tau^+}^\mu + k_{\tau^-}^\mu )\>
(\bar l \g_\mu l) = 0$ so that also the contribution of the axial-vector
operator $\langle0|\,\bar q\,\gamma^\mu\gamma_5\,b\,|B\rangle$
vanishes when contracted with the leptonic
vector current. The general form of the amplitude reads 
\beq \label{BqllAmpl}
{\cal A}^q_l  = i f_{B_q}\, m_B\, G_F\, \left[
  \left(C^q_P - {2m_l\over m_B}\, C^q_A \right) (\bar l\,\gamma_5\,l)
  + {C^q_P}'\, (\bar l\, l)\, \right]\,,
\eeq
and the corresponding branching ratio is 
  \beqa \label{BqllRate}
\BR(\Bqll) &=& f_{B_q}^{2} \tau_B \frac{m_{B}^{3}\, G_{F}^{2}}{8\,
\pi}\sqrt{1-\frac{4 m^2_l}{m^2_B}}
\left[ \left|C^q_P - \frac{2\,m_l}{m_B}\, C^q_A\right|^2
+ \left(1-\frac{4 m^2_l}{m^2_B}\right)\left|{C^q_P}'\right|^2\,\right]\,.
\eeqa
In the SM, ${C^q_P}'$ and ${C^q_P}$ arise from penguin diagrams
with physical and unphysical neutral scalar exchange, and are
suppressed as $\sim (m_b/m_W)^2$ \cite{SkKa}. The decay rate is
then determined by
\beq \label{CqASM}
\Big[C^q_A\Big]_{\rm SM} = {\alpha \,  V_{tb}^*\,V_{tq}\over
  \sqrt8\,\pi\sin^2\theta_W}\, Y_0(x_t) \,,
\eeq
where $x_t=m_t^2/m_W^2$, and at leading order  \cite{inami-lim}
\beq \label{eqY}
  Y_0(x) = {x\over8}\, \left[{x-4\over x-1}+{3x\over(x-1)^2}\,\ln x \right].
\eeq
Using the parameters listed in the Appendix, we find
$ \BR^{\rm SM}(\Bqtt) = 9.1\times10^{-7}\, \left|V_{tq}/V_{ts}\right|^2 \,$.
We stress that as for \Btn, also for this decay  theoretical predictions
are plagued by the large uncertainty in the  value of $f_{B_q}\,$,
which can easily mask new physics effects.

%%%%%%%%%%%%%%%%%%%%%%%%%%%%%%%%

\subsection{The decay \bqnn. }

The decay \bqnn\ was thoroughly studied in \cite{bsnn}.
In the presence of new physics, the flavor of
the two final state neutrinos can differ. Still,
under the only assumption that neutrinos
are purely left-handed and effectively massless,
the general form of the amplitude has the remarkably simple form \cite{bsnn}
\beqa\label{bqnnAmpl}
{\cal A}^{qij}&=& G_{F}\left\{\
C^{qij}_{L}\left( \bar{q}_{L}\g_{\m}  b_{L} \right)
\left(\bar{\nu}^{i}_{L}\g^{\m} \nu ^{j}_{L} \right) +
C^{qij}_{R}\left( \bar{q}_{R}\g_{\m}  b_{R} \right)
\left(\bar{\nu}^{i}_{L}\g^{\m} \nu ^{j}_{L} \right) \right\}.
\eeqa
In terms of the coefficients in (\ref{heff}) we have
$C_{L,R}= [(C_V-C'_V) \pm  (C_A-C'_A)]/4 $.
Summing over the undetected neutrino flavors the branching ratio normalized
to the semileptonic decay  reads
\beq\label{bqnnRate}
\BR(b\to  X_q \nu \bar\nu)  =
\frac{\sum_{ij}\left(|C^{qij}_{L}|^{2} + |C^{qij}_{R}|^{2}\right) }
{8\,|V_{cb}|^2 f_{PS}(m_{c}^{2}/m_{b}^{2})}\> \BR(b\to  X_c e \bar\nu )\,,
\eeq
where $f_{PS}(x)=1-8x+8x^3-x^4-12x^2\ln x \approx 0.5$ for
$x=m_c^2/m_b^2$ is the phase space factor for the semileptonic decay.
In the SM the  decay proceeds via $W$ box and $Z$ penguin diagrams, and
only one operator contributes to the decay:
$O^{\rm SM}_L=\left( \bar{q}_{L}\g_{\m}  b_{L} \right)
\sum_i \left(\bar{\nu}^i_{L}\g^{\m} \nu^i_{L} \right)\,$.
The corresponding coefficient reads
\beq \label{CqLSM}
\Big[C^q_L\Big]_{\rm SM} = {\sqrt2\,\alpha   V_{tb}^*\,V_{tq}
 \over \pi \sin^2\theta_W\,}\, X_0(x_t) \,,
\eeq
where  \cite{inami-lim}
\beq \label{eqX}
X_0(x) = {x\over8}\, \left[{2+x\over x-1}+{3x-6\over(x-1)^2}\,\ln x \right].
\eeq
The additional  $1/m_b^2$ and $\alpha_s$ corrections to this result
can be found  in \cite{bsnn,BuBuL,Adametal}.
In contrast to the previous decays, theoretical predictions for
\bqnn\ are remarkably free from uncertainties.
In fact all the parameters entering in (\ref{bqnnRate}) and (\ref{CqLSM})
are known with good accuracy (the main uncertainty
comes from $m_t$), there are no long distance effects
and QCD corrections are small \cite{bsnn}.
{}From the theoretical point of view,  new physics
affecting this decay  could be identified in a very clean way.
At leading order, the  SM prediction for the branching ratio is
$\BR^{\rm SM}(\bqnn) = 4.4\times 10^{-5}\, \left|V_{tq}/V_{ts}\right|^2 \,$.

%%%%%%%%%%%%%%%%%%%%%%%%%%%%%%%%%%%%%%%%%%%%%%%%%%%%

\section{The decay \bxll.}

\subsection{General operator basis.}

In the SM, the effective Hamiltonian for the weak decay
\bsll\ is defined in terms of a  set of ten
effective operators $O_1$--$O_{10}$
\cite{Witten}.\footnote{For simplicity we will restrict ourselves to the
case when the final hadronic system carries strangeness $(X=X_s)$.
Generalization to the case $X=X_d$ requires introducing the
additional operators $O^u_1=(\bar d_L^\alpha \gamma_\mu b_L^\alpha)\,
(\bar u_L^\beta \gamma_\mu u_L^\beta)\,$ and
$ O^u_2 = (\bar d_L^\alpha \gamma_\mu b_L^\beta)\,
(\bar u_L^\beta \gamma_\mu u_L^\alpha)\,$, keeping
the terms proportional to $V_{ub}$ and including
new long distance effects.}
At leading order, and neglecting small contributions induced only
through operator mixing,  the operator basis can be truncated
to the following set \cite{gsw}
\beqa \label{setSM}
O_1 &=& (\bar s_L^\alpha \gamma_\mu b_L^\alpha)\,
(\bar c_L^\beta \gamma_\mu c_L^\beta) \nnu \\
O_2 &=& (\bar s_L^\alpha \gamma_\mu b_L^\beta)\,
(\bar c_L^\beta \gamma_\mu c_L^\alpha) \nnu \\
O_7 &=&
% \frac{e}{16\pi^2} % take care of this coupling !
(\bar s^\alpha \sigma_{\mu\nu}[m_bP_R+m_qP_L] b^\alpha)F^{\mu\nu}  \nnu \\
O_9 &=&(\bar s_L^\alpha \gamma_\mu b_L^\alpha)\,(\bar l\gamma_\mu l)\nnu \\
O_{10} &=& (\bar s_L^\alpha \gamma_\mu b_L^\alpha)\,
(\bar l \gamma_\mu\gamma_5 l )\,.
\eeqa
In the models we want to study,  a larger set of non-renormalizable
operators arises. As it will become clear in the next section, after
Fierz transformation
squark exchange induces {\it at the tree level}
the new operators $O'_2\,$, $O'_9\,$, $O'_{10}\,$ which are
analogous to the corresponding  operators in (\ref{setSM})
with the replacement
$(\bar s_L \gamma_\mu b_L) \to (\bar s_R \gamma_\mu b_R) $.
While  $O'_9\,$ and $O'_{10}\,$ contribute directly to the decay,
$O'_2\,$ enters only at the one loop level, so that the corresponding short
distance contribution is small. However, $O'_2\,$ induces
also new long distance
effects associated with  $\bar c c$ resonances which can further enhance the
decay rate above the SM. For this reason  we  include $O'_2\,$ in our set.
At the new physics scale $\tilde m \gsim 100\,$GeV where these operators are
generated, other operators from new physics appear only  at the loop level.
Since they give only suppressed short distance
contributions we set to zero the corresponding high energy coefficients.
However, in the evolution from the scale $\tilde m$
down to $m_b$ operator mixing occurs, and from the point of view
of the low energy theory a clear distinction between tree-level
and loop-level contributions is lost. This forces us to extend
the basis to the following set
\beqa \label{setRp1}
O'_1 &=& (\bar s_R^\alpha \gamma_\mu b_R^\alpha)\,
(\bar c_L^\beta \gamma_\mu c_L^\beta)  \nnu \\
O'_2 &=& (\bar s_R^\alpha \gamma_\mu b_R^\beta)\,
(\bar c_L^\beta \gamma_\mu c_L^\alpha)  \nnu \\
O'_7 &=&
(\bar s^\alpha \sigma_{\mu\nu}[m_bP_L+m_sP_R] b^\alpha)F^{\mu\nu}  \nnu \\
O'_9 &=& (\bar s_R^\alpha \gamma_\mu b_R^\alpha)\,
(\bar l \gamma_\mu l )  \nnu \\
O'_{10} &=& (\bar s_R^\alpha \gamma_\mu b_R^\alpha)\,
(\bar l \gamma_\mu\gamma_5 l )\,.
\eeqa
In addition to the new set $\{O'_i\}\,$, the exchange of
sleptons induces at the tree level
new scalar operators  which also contribute directly to the decay
\beqa \label{setRp2}
O^S_9 &=& (\bar s_R^\alpha b_L^\alpha)\,(\bar l  l )  \nnu \\
O^S_{10} &=& (\bar s_R^\alpha  b_L^\alpha)\,(\bar l \gamma_5 l )  \nnu \\
{O^S_9}' &=& (\bar s_L^\alpha b_R^\alpha)\,(\bar l  l ) \nnu \\
{O^S_{10}}' &=& (\bar s_L^\alpha  b_R^\alpha)\,(\bar l \gamma_5 l )\,.
\eeqa
These operators do not mix with $\{O_i\}$ and $\{O'_i\}\,$, neither
with new loop-induced four-quarks scalar operators, which at lowest
order do not contribute to the decay.
Hence we can truncate the basis to the subset of operators
listed in  (\ref{setSM}), (\ref{setRp1}) and~(\ref{setRp2}).

Including the new physics, we can now write the Hamiltonian density as
\beqa \label{heffbstt}
% -\frac{1}{G_F}\,
{\cal H}_{\rm eff} & = & - G_F \Bigg\{
\Big(\bar{s}\,\g_{\m}\, (C_{9}  P_L + C'_{9}  P_R)\, b\, \Big)
\left(\bar{l}\g^{\m}l\right) +
\Big(\bar{s}\, \g_{\m} \, (C_{10} P_L + C'_{10} P_R)\,  b\, \Big)
\left(\bar{l}\g^{\m}\g^5 l \right)
\nnu \\
& - &
2\,i\,\frac{q^{\n}}{q^2}\,\left(\bar{s}\,\sigma_{\m\n}\,
\Big[\,( C_{7} P_R + C'_{7} P_L ) m_b +
       ( C_{7} P_L + C'_{7} P_R )\, m_s\,\Big]\, b\, \right)
\left( \bar{l}\g^{\m} l \right)
\nnu\\
& + &
\Big(\,\bar{s}\,(C^{S}_{9}  P_L + C^{\prime S}_{9}  P_R) \,b\,\Big)
\left( \bar{l} l \right) +
\Big(\,\bar{s}\, (C^{S}_{10}  P_L + C^{\prime S}_{10}  P_R)\,b\,\Big)
\left( \bar{l}\g_5 l\right)\,\Bigg\}\,,
\eeqa
where we  have factored out the Fermi constant $G_F$ 
so that the coefficients $C$ are dimensionless.

The SM contributions to (\ref{heffbstt}) enter through the
coefficients
\beq \label{ci}
C_{i}  =  \frac{\a}{\sqrt{2}\p}\,V_{tb}^{}V_{ts}^*\,c_{i}
\quad (i=7,10)\,, \qquad {\rm and} \qquad
C_{9}  =  \frac{\a}{\sqrt{2}\p}\,V_{tb}^{}V_{ts}^*\,c_{9}^{eff}\,,
\eeq
where $c_{i}$ are the usual QCD improved Wilson coefficients \cite{gsw}.
In the leading logarithmic approximation, and neglecting
small operator mixings, we have \cite{gsw,bm,AliGM}
\beqa \label{coeffSM}
c_1(m_b) &=& \frac{1}{2}
\left[\eta^{\frac{6}{23}}-\eta^{-\frac{12}{23}}\right]c_2(M_W)  \nnu \\
c_2(m_b) &=& \frac{1}{2}
\left[\eta^{\frac{6}{23}}+\eta^{-\frac{12}{23}}\right]c_2(M_W)   \nnu \\
c_7(m_b) &=& \eta^{\frac{16}{23}} \left\{
c_7(M_W) - \left[  \frac{58}{135} (\eta^{-\frac{10}{23}}-1) +
\frac{29}{189} (\eta^{-\frac{28}{23}}-1) \right] c_2(M_W) \right\}  \nnu \\
c_9(m_b) &=& c_9(M_W) - \frac{4 \pi}{\alpha_s(M_W)} \left[
           \frac{4}{33} ( 1 - \eta^{\frac{11}{23}})  -
\frac{8}{87} ( 1 - \eta^{\frac{29}{23}}) \right] c_2(M_W)  \nnu \\
c_{10}(m_b) &=& c_{10}(M_W)
\eeqa
where $ \eta = \alpha_s(M_W)/\alpha_s(m_b) \,$ and
at the renormalization point $\mu \sim M_W\,$,
$c_1(M_W)=0$ and  $c_2(M_W)=1$. The remaining three coefficients
are functions of $m_t/M_W$ and their explicit expressions can be
found in \cite{gsw}. Using $m_t=176\,$GeV and $\sin^2\theta=0.23$
(where $\theta$ is the weak mixing angle) we obtain
$c_7(M_W)=-0.195\,$, $c_9(M_W)=2.056\,$ and $c_{10}(M_W)=-4.415\,$
(with $c_9(M_W)$ defined according to the prescription
given in \cite{gsw}).
The coefficient $c_9^{eff}$ appearing in (\ref{ci}) includes
the $s$ dependence
induced by the one-loop matrix element of the four-quark operators, and
reads \cite{gsw}
\beq \label{c9eff}
c_9^{eff} = c_9(m_b) + \big[3 c_1(m_b) + c_2(m_b)\big]\> g(\mch,\sh).
\eeq
where $\mch=m_c/m_b$ and $\sh=s/m^2_b\,$.
In our numerical analysis, we use the expression for
$g(\mch,\sh)$ given in \cite{bm}. An additional contribution to this
decay mode  comes from the long distance effects
associated with on-shell and off-shell $c\bar c$
resonances. There are six known resonances that can
contribute. They generate an additional term which has the same
structure as $O_9$. Hence, it is convenient to include the resonance
contributions directly into $c_9^{eff}$
by making the replacement \cite{longdist}
\beq \label{resonances}
g(\mch,\sh)  \to \tilde g(\mch,\sh) =
g(\mch,\sh) - \frac{3\pi}{\a^2}\sum_{V=J/\psi , \psi',\dots }
\frac{\kappa_V \hat{M_V} {\hat\Gamma}(V\to l^+l^-)}
{\sh-\mvh^2+i\mvh\hat{\Gamma}_V}\,,
\eeq
where $\hat{\Gamma}_V$ and $\hat{\Gamma}(V\to l^+l^-)$ are respectively the
total and partial decay width of the vector meson resonances normalized
to $m_b\,$, while $\mvh$ is the normalized vector meson mass.
The  values of the masses, widths and leptonic branching ratios
of the six $\bar c c$ resonances  can be found in \cite{pdg},
while the phenomenological factor $\kappa_V\sim 2.3$
is calculated by fitting the $B\to J/\psi K^*$ amplitude
to the experimental rate \cite{jpsi}.
The replacement (\ref{resonances}) to model the long distance contributions
from off-shell resonances is not a rigorous procedure \cite{LDnew}.
Here we adopt  this simple prescription in order to compare the
long distance effects with the new physics 
short distance contributions.

The low energy coefficients of the new operators in (\ref{setRp1})
can be determined in the same way as the SM coefficients.
Due to the vectorlike nature of QCD, the matrix of anomalous dimensions
for the set $\{O'_i\}$ is the same as for the standard basis.
Moreover, in the models that we will study 
in the next section, operators induced by squark exchange at the scale
$m_{\tilde q}\,$ do not renormalize. Therefore,
to  take into account the  QCD effects on the new physics
operators, the only new scale that needs to be
introduced is $\tilde m = m_{\tilde l}\,$.
We fix $\tilde m = 100\,$GeV.
We note that in this case the renormalization group evolution is still
controlled by the  QCD $\beta$-function with five active flavors.
Modifications to account for the case $m_{\tilde l} > m_t$
are straightforward. In the models discussed in the next section,
$C'_1(\tilde m)=0$.
Since new physics generates $O'_7$ only at the loop level, we will also set
$C'_7(\tilde m)=0$ and $C'_7(m_b)$ arises only from operator mixing.
Then the set of equations that determines the
low energy coefficients $C'_i(m_b)$ is the same than (\ref{coeffSM})
with the replacements $c_7(M_W)  \to C'_7(\tilde m)=0$
and $\eta \to \eta' = \alpha_s(\tilde m)/\alpha_s(m_b)\,$.
In particular, we have included in ${C_9}'(m_b)$
also the additional long distance contributions induced by
$O'_1$ and $O'_2\,$ in the same way as for $c_9^{eff}$ in
(\ref{c9eff}).

Finally, the evolution of the coefficients of the scalar
operators in (\ref{setRp2}) is controlled by the anomalous
dimensions  of $O^S_9\,$, $O^S_{10}\,$, ${O^S_9}'\,$ and ${O^S_{10}}'\,$:
$\gamma_{O^S}=-4$ \cite{anomsdim}.

\subsection{Inclusive rate and various observables.}

\bigskip
Neglecting non-perturbative ($\sim 1/m_b^2$)  corrections \cite{fls} the
inclusive \bsll\ decay width as a function of the invariant mass of
the lepton pair $q^2 = m^2_{l^+ l^-}$ is given by
\beq \label{rate}
\frac{d\G(\sh)}{d\sh} = \frac{G_{F}^{2} m_b^5}{384 \p^3}\,
\l^{1/2}(1,\sh,\msh^2) \,\sqrt{1-\frac{4 \mlh^2}{\sh}}\,\Sigma(\sh)\ ,
\eeq
where  $\sh = q^2/m_b^2$, $\mh_i = m_i/m_b\,$,
$\l (a,b,c) = a^2 + b^2 + c^2 - 2 (a b + b c + a c) \,$
and
\beqa \label{sigma}
\Sigma(\sh) &=&
4\, \left(1+\frac{2\mlh^2}{\sh}\right)\,
\Bigg[
\frac{1}{\sh}\left(|C_{7}|^2+|C'_{7}|^2 \right)F_1(\sh,\msh^2)
+
\frac{4\msh}{\sh}\Re\left(C^*_{7}C'_{7}\right) F_2(\sh,\msh^2)
\nnu\\ &+&
3\,\Re\left(C^*_{7} C_{9}+C^{\prime *}_{7}C^{\prime}_{9}
\right)F_3(\sh,\msh^2)
- 3\,\msh \sh \, \Re\left(2C^*_{7}C^{\prime}_{9}+
2C^*_{9}C^{\prime}_{7} + C^*_{9}C^{\prime}_{9}\right)
\Bigg]
\nnu\\ &+&
\left(|C_{9}|^2+|C^{\prime}_{9}|^2 +|C_{10}|^2+|C^{\prime}_{10}|^2\right)
 F_4(\sh,\msh^2,\mlh^2) -
12 \msh \sh
\left(1-\frac{6\mlh^2}{\sh}\right)\Re(C^*_{10}C^{\prime}_{10})
\nnu\\ &+&
6\, \mlh^2\,
\left(|C_{9}|^2+|C^{\prime}_{9}|^2-|C_{10}|^2-|C^{\prime}_{10}|^2 \right)
F_5(\sh,\msh^2)
\nnu\\ &+&
\frac{3}{2} \sh\, \left[ \left(1-\frac{4\mlh^2}{\sh}\right)
\left( |C^{S}_{9}|^2+|C^{\prime S}_{9}|^2 \right) +
\left( |C^{S}_{10}|^2+|C^{\prime S}_{10}|^2 \right)
\right]\,F_5(\sh,\msh^2)
\nnu\\ &+&
6\, \sh \, \msh \, \left[ \left(1-\frac{4\mlh^2}{\sh}\right)
\Re\Big({C^S_9}^*\,C^{\prime S}_{9}\Big)+
\Re\Big({C^S_{10}}^*\,C^{\prime S}_{10}\Big) \right]
\nnu\\ &+&
6\,\mlh  \Re\Big({C^{\prime S}_{10}}^*\,C_{10}+
{C^{ S}_{10}}^*\,C^{\prime}_{10}\Big)\,
\left(F_5(\sh,\msh^2) - 2\,\msh^2\right)
\nnu\\ &-&
6 \,\mlh\,\msh \, \Re\Big({C^{\prime S}_{10}}^*\,C^{\prime }_{10}+
{C^S_{10}}^*\,C_{10}\Big)\,
\left(F_5(\sh,\msh^2) -2 \right)  \,.
\eeqa
The functions $F_i$  read
\beqa \label{kinfuncs}
F_1(\sh,\msh^2)&=& 2 (1+\msh^2)\, (1-\msh^2)^2 -\sh (1 + 14 \msh^2
+\msh^4)-\sh^2 (1+\msh^2)\,,\nnu\\
F_2(\sh,\msh^2)&=& 2 (1-\msh^2)^2  -\sh (4+4\msh^2+\sh)\,,\nnu \\
F_3(\sh,\msh^2)& =&(1-\msh^2)^2 - \sh(1+\msh^2)\, ,  \nnu\\
F_4(\sh,\msh^2,\mlh^2) &=& (1-\msh^2)^2 +\sh (1+\msh^2) -2 \sh^2 +
\l(1,\sh,\msh^2) \frac{2\mlh^2}{\sh}\,,\nnu \\
F_5(\sh,\msh^2)&=& 1- \sh +\msh^2\,.
\eeqa
The forward-backward asymmetry  $A_{\rm FB}$  \cite{amm} is defined
with respect to the angular variable $c_\theta=\cos\theta$ where
$\theta$ is the angle of the lepton $l^-$ with respect to the
$b$-direction in the $l^+ l^-$ center-of-mass system: 
\beq \label{Afb}
\fba \equiv \frac{1}{d\Gamma(\sh)/d\sh}
\left[
\int_0^1\,dc_\theta \frac{d^2 }{d\sh\, d c_\theta}\Gamma(\sh,c_\theta) -
\int_{-1}^0\,dc_\theta \frac{d^2 }{d\sh\, d c_\theta}\Gamma(\sh,c_\theta)
\right]\,.
\eeq
We obtain :
\beq \label{asymmetry}
\fba = 3\, \l^{1/2}(1,\sh,\msh^2) \,\sqrt{1-\frac{4 \mlh^2}{\sh}}
\frac{\Delta(\sh)}{\Sigma(\sh)}\,,
\eeq
where
\beqa \label{delta}
\Delta(\sh) &=& \sh\,\Re\Big(C^*_9 C_{10}-C'^*_9 C^{\prime}_{10}\Big) +
2(1+\msh^2)\Re\Big(C^*_{7} C_{10}-C'^*_{7}C^{\prime}_{10}\Big) \nnu \\
&-&
4 \msh\Re\Big(C^*_7 C^\prime_{10}-C^*_{10}C^\prime_7\Big) +
\mlh\, \Re \left[ \Big( {C^S_9}^* + \msh \,  {C^{\prime S}_9}^*\Big)\,
\Big(2 C^\prime_7+C^\prime_9\Big) \right]  \nnu \\
&+&
\mlh\,
\Re \left[\Big( {C^{\prime S}_9}^* + \msh \,  {C^S_9}^*\Big) \,
\Big( 2 C_7+C_9 \Big) \right] \,.
\eeqa

Recently, polarization  measurements of final state $\tau$ leptons
in the decay \bstt\
have been proposed as a useful tool in discerning physics beyond the
SM \cite{joanne,KruSehg}.
Experimental studies of the $\tau$ polarization have been carried
out by the four LEP collaborations, and
from the analysis of the distributions of the $\tau$ decay products
in different decay modes,  the $\tau$ polarization was determined
with an error of about $10\%\,$\cite{taupol}.
For $B$-factory experiments the task of reconstructing
the $\tau$ polarization appears more challenging.
For example, in contrast to the LEP environment,
the energy of the decaying $\tau$ is not a-priori known.
However, if a comparable sensitivity can be reached,
then a measurement of the $\tau$ polarization
components could provide  crucial informations for distinguishing
among different  sources of new physics.
Following Ref.~\cite{KruSehg} we define the inclusive lepton polarization
components by introducing three orthogonal unit vectors
\beq\label{unitvec}
\eL= \frac{\Vec{p}_-}{|\Vec{p}_-|}\ , \qquad\qquad
\eN=\frac{\Vec{p}_s\times\Vec{p}_-}{|\Vec{p}_s\times\Vec{p}_-|}\ ,
\qquad \qquad \eT=  \eN\times \eL \,,
\eeq
where $\Vec{p}_-$ and $\Vec{p}_s$ are the three-momenta of $l^-$
and of the $s$ quark in the c.m.~frame of the $l^+ l^-$ system.
The differential decay rate for $b\to X_s \, l^+ l^-$ for any given spin
direction $\Vec{n}$ of the lepton $l^-$ ($\Vec{n}$ being a unit vector
in the $l^-$ rest frame) can  be written as
\beq\label{spinrate}
\frac{d\G(\sh,\Vec{n})}{d\sh} =
\frac{1}{2} \left(\frac{d\G(\sh)}{d\sh}\right)_{\rm{unpol} }\Bigg[1 +
\Big(\lop(\sh)\, \eL + \nop(\sh)\,\eN + \trp(\sh)\, \eT \Big)
\cdot\Vec{n}\Bigg]\,,
\eeq
where  $\lop$, $\nop$ and $\trp$  give the longitudinal, normal
and transverse components of the $l^-$ polarization 
as a function of $\sh\,$.
The polarization asymmetries ${\cal P}_i(\sh)$ ($i={\rm L, N, T}$)
are obtained by evaluating
\beq\label{pola}
{\cal P}_i(\sh) = \frac{d\G(\Vec{e}_i,\sh)/d\sh -d\G(- \Vec{e}_i,\sh)/d\sh}
{ d\G(\Vec{e}_i,\sh)/d\sh + d\G(-\Vec{e}_i,\sh)/d\sh}\
\eeq
for  $\Vec{e}_i=\eL\,,\eN\,,\eT\,$.
For the general form of the interaction (\ref{heffbstt}) we obtain
\beqa \label{polaL}
         \lop(\sh) &=& \frac{1}{\Sigma(\sh) }\sqrt{1-\frac{4\mlh^2}{\sh}}
\> \Bigg\{ 12\,\Re\left(C^{*}_{7}C_{10}+C^{\prime *}_{7}C^{\prime}_{10}\right)
 \left[(1-\msh^2)^2-\sh(1+\msh^2)\right]
\nnu\\  &+&
2\, \Re\left(C^{*}_{9}C_{10}+C^{\prime *}_{9}C^{\prime}_{10}\right)
\left[(1-\msh^2)^2 + \sh (1+\msh^2) - 2\sh^2\right]
\nnu \\ &-&
12\,\msh \,\sh\, \Re\left(2\,C^{*}_{7}C^{\prime}_{10}+
2\,C^{\prime *}_{7}C_{10}
+C^{*}_{9}C^{\prime}_{10}+C^{\prime *}_{9}C_{10}\right) 
\nnu \\ &-&
3\sh\left[(1-\sh+\msh^2) \,\Re\left({C^{S}_{9}}^{*}
C^{S}_{10}+{C_{9}^{\prime S}}^{*}C^{\prime S}_{10}\right)
-2\msh\,
\Re\left({C^{S}_{9}}^{*}
C^{\prime S}_{10}+{C_{9}^{\prime S}}^{*}C^{S}_{10}\right) 
\right]
\nnu \\ &-&
6\,\mlh \,
(1-\sh-\msh^2) \Re\left({C^{S}_{9}}^{*}
C^{\prime }_{10}+{C_{9}^{\prime S}}^{*}C_{10}\right) 
\nnu \\ &-&
6\,\mlh \,\msh\, 
(1+\sh-\msh^2)\, \Re\left({C^{S}_{9}}^{*}
C_{10}+{C_{9}^{\prime S}}^{*}C^{\prime}_{10}\right) 
\Bigg\}\,,
 \eeqa
\beqa                                                \label{polaN}
         \nop(\sh) &=& \frac{1}{\Sigma(\sh)}\frac{3\p }{4 }\,
\l^{1/2}(1,\sh,\msh^2)\, \sqrt{1-\frac{4\mlh^2}{\sh}}
\> \Bigg\{\frac{4\,\mlh}{\sqrt{\sh}} \Big[ 2\,
\Im(-C^{*}_{7}C^{\prime }_{10} + C^{\prime *}_{7}C_{10}) \, \msh
          \nnu \\ &+&
\Im(C^{*}_{7}C_{10}- C^{\prime *}_{7}C^{\prime }_{10}) \,(1+\msh^2)\Big] +
2\,\mlh \, \sqrt{\sh} \, \Im\left(C^{*}_{9}C_{10}
-C^{\prime *}_{9}C^{\prime }_{10}\right)
              \nnu \\ &-&
\sqrt{\sh}\,\Im\left({C^{ \prime S}_{9}}^{*}\left(2\,C_{7}+C_{9}\right)+
     {C^{  S}_{9}}^{*} \left(2\,C^{\prime}_{7}+C^{\prime}_{9}\right)+
\left({C^{\prime S }_{10}}^{*}C_{10}+{C^{ S }_{10}}^{*}C^{\prime}_{10}\right)
\right)
            \nnu \\  &-&
  \msh\,\sqrt{\sh}\,
\Im\left({C^{ S}_{9}}^{*}\left(2\,C_{7}+C_{9}\right)+
     {C^{\prime  S}_{9}}^{*} \left(2\,C^{\prime}_{7}+C^{\prime}_{9}\right)+
\left({C^{\prime S }_{10}}^{*}C^{\prime}_{10}+{C^{ S }_{10}}^{*}C_{10}\right)
\right)\Bigg\}\,,
\eeqa
\beqa                                               \label{polaT}
      \trp(\sh)&=&\frac{1}{\Sigma(\sh)}\frac{3\p }
{2\sqrt{\sh}}\l^{1/2}(1,\sh,\msh^2)
\> \Bigg\{\,
\mlh\Bigg[-\frac{4}{\sh}\, \left(|C_{7}|^2 -|C^{\prime}_{7}|^{2}\right)
\,(1-\msh^2)^2
 \nnu  \\ &+&
\Re\left(C^{*}_{10}C_{9} +C^{\prime *}_{10}C^{\prime}_{9}\right)(1-\msh^2)
-8\, \msh \Re\left(C^{\prime
*}_{7}C_{9}-C^{*}_{7}C^{\prime}_{9}\right) -\sh
\left(|C_{9}|^2-|C^{\prime}_{9}|^2\right)
        \nnu \\ &-&
4\,\Re\left(C^{*}_{7}C_{9} -C^{\prime
*}_{7}C^{\prime}_{9}\right)(1+\msh^2) + 2 \,\Re
\left(C^{*}_{7}C_{10} +C^{\prime
*}_{7}C^{\prime}_{10}\right)\, (1-\msh^2)   \Bigg]
\nnu \\  &+&
\frac{\sh}{2}\Bigg[\left(1-\frac{4\mlh^2}{\sh}\right)
\left[
\Re\left({C^{ \prime S}_{9}}^{*}
C_{10}+{C^{S}_{9}}^{*}C^{\prime}_{10}\right)+
\msh\,\Re\left({C^{S}_{9}}^{*}
C_{10}+{C^{\prime S}_{9}}^{*}C^{\prime}_{10}\right)
\right]
\nnu \\ &+&
\Re\left({C^{\prime S}_{10}}^{*}
\left(2\,C_{7}+C_{9}\right)+
     {C^{S}_{10}}^{*} \left(2\,C^{\prime}_{7}+
     C^{\prime}_{9}\right)\right)
\nnu \\ &+&
\msh \,\Re\left({C^{S}_{10}}^{*}
\left(2\,C_{7}+C_{9}\right)+
     {C^{\prime S}_{10}}^{*} \left(2\,C^{\prime}_{7}+
     C^{\prime}_{9}\right)\right)
\Bigg]
\Bigg\} .
\eeqa

The possibility of measuring the longitudinal polarization of the $\tau$
in the decay \bstt\  was first proposed in \cite{joanne}.
This analysis was extended to include the other two polarization
components  $\nop$  and $\trp$  in  \cite{KruSehg}.
The SM results are recovered from our formul\ae\
by setting   the $C^\prime_i$ and the
$C^{(')S}_{9,10} $ coefficients to zero. In this limit our expressions
for $\lop(\sh)$ and $\nop(\sh)$ agree with the results given
by Kr\"uger and Sehgal \cite{KruSehg}, and when we  set $m_s=0$
 in (\ref{polaL}) we find agreement with  $\lop$ as given in  \cite{joanne}.
However,  the SM limit of $\trp$ disagrees with eq.~(5.5) 
in \cite{KruSehg} for the factor of two multiplying the term   
$\Re(C_{7}^{*} C_{10}) $ in the third line of (\ref{polaT}).
% Since the polarization components $\lop$, $\trp$ and $\nop$ involve
% different combinations of the effective couplings
% with respect to the inclusive branching ratio and
% to $A_{\rm FB}$, they provide independent informations.

To  derive numerical results for the inclusive branching ratios
and for the averaged values of the asymmetries,
we minimize long distance effects
by imposing cuts on the dilepton invariant mass.
For the muon channel we select the region below  the
resonances $\sh < 0.4\,$,  while for the tau channel we require
$\sh > 0.6\,$,  above the $\psi'\,$. The remaining effects of the four
additional  $\bar c c$ resonances in the tail of the invariant
mass distribution for \bstt\ are not very large.

At leading order, the SM results for the various quantities are as follows.
The SM inclusive branching ratio for muons is predicted to be
${\BR}(b\to X_s \, \m^+ \m^-)_{\scriptscriptstyle \hat s < 0.4}
= 4.3\times 10^{-6}\,$. The purely short distance
contribution yields in the same  region 
${\BR}^{\rm sd}(b\to X_s \, \m^+ \m^-)_{\scriptscriptstyle \hat s < 0.4}
= 3.9\times 10^{-6}\,$.
Due to phase space suppression and  to the
different cut, the tau  production rate is more than one order of
magnitude smaller. We find
${\BR}(b\to X_s \, \t^+ \t^-)_{\scriptscriptstyle \hat s > 0.6}
= 1.5\times 10^{-7}$ and a similar number  for the purely
short distance contribution.
Below our cut, the average value of the $\mu$  forward-backward asymmetry
is rather small $\left\langle A^\mu_{\rm FB}
\right\rangle_{\scriptscriptstyle \hat s < 0.4} = -0.01\,$.
This is due to the fact that the asymmetry changes sign in this region,
and in taking the average large cancellations occur.
For the $\tau$ lepton the  asymmetry is larger $\left\langle A^\tau_{\rm FB}
\right\rangle_{\scriptscriptstyle \hat s > 0.6} = -0.13\,$.
Only the longitudinal polarization
asymmetry  $\lop$ is  significant  for the $\mu\,$:   
$\left\langle \lop^\mu
\right\rangle_{\scriptscriptstyle \hat s < 0.4} = -0.57$
while all the three components are sizeable for $\tau\,$.
We find the following average values
$\left\langle \lop^\tau \right\rangle_{\scriptscriptstyle \hat s > 0.6}
= -0.34\,$,
$\left\langle \trp^\tau \right\rangle_{\scriptscriptstyle \hat s > 0.6}
= -0.40\,$,
$\left\langle \nop^\tau \right\rangle_{\scriptscriptstyle \hat s > 0.6}
= 0.05\,$.

Notice that the normal  component
$\nop$, though small, is a T-odd quantity and it is
considerably larger than the corresponding normal
polarization of leptons in $K_L\to\pi^+\m^-\bar{\n}$ or
$K^+\to\pi^+\m^+\m^-$ \cite{kdecay}.
However, this should not be interpreted as signaling  $CP$ violation.
In fact, since we assume real couplings, the leading contribution comes
from  the absorptive part of the effective
coupling $C_9$ (and $C'_9\,$ in the new physics analysis),
which are dominated by the $\bar c c$ real intermediate states
(cf.(\ref{resonances})).
It is in principle possible to remove this background to
$CP$ violating effects by measuring the difference between
the $\nop$ components for example in $B^+$ and $B^-$ decays. This is
because $CP$ violating phases yield asymmetries with the same sign,
while phases originating from strong interactions cancel
off \cite{CPphase}.

\section{SUSY without R-parity }
\bigskip

In this section we present a short introduction to SUSY models
where R-parity is not imposed, and $L$ is (mildly) violated
already at the renormalizable level. Next we will embed these
models in the framework of an Abelian horizontal symmetry
\cite{horizontal} that will provide us with the additional
theoretical constraints needed to derive numerical predictions.
As it was discussed  in
\cite{BGNN1}, this procedure has the additional advantage of
ensuring that in the particular models we will discuss, the
present constraints on the R-parity violating couplings are
satisfied.

\subsection{R-parity violating couplings }

The field content of the SM together with the
requirement of $SU(2)_L\times U(1)_Y$ gauge invariance, implies that
at the renormalizable level the most general Lagrangian possesses
additional accidental $U(1)$ symmetries, corresponding to conserved
baryon and lepton flavor ($L_i$) quantum numbers. The
conservation of $B\,$, $L_i\,$ and hence of total lepton number
($L=\sum_i L_i$) naturally explains nucleon stability as well as the
non observation of $L$ and $L_i$ violating  transitions.
In SUSY extensions of the SM, additional gauge and
Lorentz invariant  terms are allowed, which violate $B$, $L_i$ and $L\,$.
Denoting collectively  by $\hat H_\alpha$ ($\alpha=0,1,2,3$) the
supermultiplets containing the down-type Higgs and the left-handed lepton
doublets, which transform in the same way under the gauge group, the
following $L_i$ and $L$ violating superpotential  terms arise
\beq \label{Lviol}
W_{\Lbs} =  \mu_\alpha \hat H_\alpha \hat  H_u  +
\la_{\alpha\beta k} \> \hat H_\alpha \, \hat H_\beta \,  \hat { l^c_k} +
\lp_{\alpha   j  k} \> \hat H_\alpha  \,\hat Q_j\, {\hat d^c_k}\,.
\eeq
Here $\hat Q_i $ and $ \hat { d^c_i} $ denote the quark doublet and
down-quark singlet superfields, $\hat { l^c_i}$ are the
lepton singlets and  $\hat  H_u$ contains the up-type Higgs field.
There are also  renormalizable terms which violate  $B$,
$ W_{\Bbs} = \lambda^{\prime\prime}_{ijk}\,
\hat u^c_i\,\hat d^c_j\,\hat d^c_k \,, $
and physics at some large scale $M_\Lambda $ can induce
additional dimension 5 $B$ and $L$ violating  terms like
$({\Gamma^\prime_{\alpha ijk} / M_\Lambda})
\, \hat H_\alpha \, \hat Q_i \, \hat Q_j \, \hat Q_k + \dots \,$.

To forbid the dangerous dimension 4 terms, a parity quantum number
${\rm R}=(-1)^{3B + L + 2S}$ ($S$ being the spin) is
assigned to each component field, and invariance under R
transformations is imposed. However, even if suppressed by the Plank mass,
the R-parity conserving dimension 5
terms can still induce too fast proton decay
unless $\Gamma^\prime \lsim 10^{-8}$ etc.
{}From a phenomenological
point of view, the first priority is to ensure the absence of
operators leading to fast nucleon decay, and in this respect other
discrete symmetries can be more effective than R \cite{Ross}.
These interesting alternatives forbid
dimension 4 and 5 $B$ violating terms but do not imply  the same for
the $L$ non-conserving terms.  Since a mild
violation of $L$ can be phenomenologically tolerated, SUSY extensions
of the SM with highly suppressed $B$ violation but without R parity
and without $L$ number, represent interesting alternatives to the
Minimal Supersymmetric Standard Model (MSSM).
We henceforth assume that $B$ is effectively conserved, and we
concentrate only on the $L$-violating terms contained  in (\ref{Lviol}).

The first term in (\ref{Lviol}) can mix the fermions with the Higgsinos,
resulting in too large neutrino masses \cite{BGNN1,HallSuzuki,Rnumass,RGnu}.
In  Ref.~\cite{BGNN1} this problem was solved by assuming that the 
down-type Higgs transforms differently from the lepton doublets under a 
horizontal
symmetry. The different charge assignments  can generate enough suppression
of the lepton mixing with $H_u\,$.
However, this does not occur in the models  discussed in the next section
(and in Ref.~\cite{YossiYuval}). In fact in these models the down-type Higgs
and the $\tau$ doublets have the same horizontal charge, and in the model
labeled below as Model II, this is  true also for the muon doublet.
We can still avoid generating too large neutrino masses under the
assumption that the  soft SUSY breaking terms are universal.
As it is discussed in \cite{RGnu}, this assumption implies the following:
\vspace{-8pt}
\begin{itemize}\itemsep=-4pt
\item[($i$)]
Modulo small violation of the universality conditions at the electroweak
scale induced by the renormalization group running of the soft SUSY breaking
parameters, the combination 
$H_d \equiv \mu_\alpha H_\alpha /\sqrt{\mu_\alpha \mu_\alpha}$ 
corresponds to the  down-type Higgs. Namely  its scalar component is
the only  other field, besides $H_u$, acquiring a non-vanishing vacuum
expectation value.
\item[($ii$)]
In the basis $\{\hat H_d,\hat L_i\}$
(where $\hat L_i$ with $i=1,2,3$ denote the three combinations orthogonal to
$\hat H_d$) the couplings  $\la_{0ij}$ and $\lp_{0ij}$ are respectively
the Yukawa couplings  $Y^l_{ij}$ of the leptons and $Y^d_{ij}$ of the
down-type quarks.
\end{itemize} \vspace{-8pt}
As it will become clear in the next subsection, this short discussion is
relevant only for the subset of the $\hat H_\alpha$ multiplets that carry
the same minimum charge.
For  doublets having horizontal charge assignments different from  the
minimum one (as for example  for the electron doublet)  the horizontal
symmetry by itself defines  to a 
very good approximation the physical fields.

Once the fields are rotated to the physical basis, the $L$ violating
trilinear terms contained in (\ref{Lviol}) read
\beq \label{Lviolmass}
\la_{i j k} \> \hat L_i\, \hat L_j\,  \hat {l^c_k} +
\lp_{i j k} \> \hat L_i\,\hat Q_j\,\hat {d^c_k}\,,
\eeq
where $\la_{i j k}=-\la_{j i k} $ due to the antisymmetry
in the $SU(2)$ indices.
Several of the $\la$ and $\lp$ couplings are strongly
constrained by the existing phenomenology \cite{susynoR}.
The best limits are for couplings involving
fermions of the first two generations ($i,j,k=1,2$) while for couplings
involving more than a single third generation field
the existing limits are much weaker, and in some cases no bounds exist
to date. This situation is interesting since in general 
models that can explain the observed fermion mass hierarchy also
predict that R-parity violating couplings involving 
third generation fields are the largest ones. 

\subsection{ R-parity violation
                    in the framework of horizontal symmetries}
\vspace{0.1in}

In order to evaluate  the effects of the new
R-parity violating interactions, we need to estimate quantitatively
the coefficients $\lambda$ and $\lambda^\prime$.
We work in the framework of the supersymmetric  models
with horizontal symmetries that have been
thoroughly investigated in \cite{horizontal}.
These models  successfully  predict  the order of magnitude
of the fermion masses and CKM  angles, and can
also explain the suppression of $L$ violation \cite{YossiYuval}
and $B$ violation \cite{Valery} in SUSY models without
R-parity. Assuming a 
horizontal $U(1)$ symmetry allows us to estimate the size of the  $L$
violating couplings, and to work out numerical predictions for various
observables measurable in $B$ decays.

In  the models  we are interested in, there are no additional  fields
in the low energy spectrum  with respect to minimal SUSY.
However, a  charge $H(\hat\psi)$ of an Abelian  horizontal symmetry 
${\cal{H}} = U(1)_{H}$ is assigned to 
each supermultiplet $\hat \psi\,$.  
${\cal{H}}$ is explicitly broken by a small
parameter $\varepsilon$ with charge $H(\varepsilon)=-1\,$
giving rise to a set of selection rules for the effective couplings
of the low  energy  Lagrangian \cite{horizontal}.
If we assume that each of the lepton, quark and Higgs superfields
carries positive or zero  charge, the selection rule
relevant for the present  discussion is that the effective coupling
$g_{abc}$ for a general trilinear superpotential term
$\hat \psi_a\hat \psi_b\hat \psi_c$ is of order
$g_{abc}\sim \varepsilon^{H(\hat\psi_a)+H(\hat\psi_b)+H(\hat\psi_c)}\,$.
Therefore the leptons and down-type quarks Yukawa couplings are
respectively  of order
$Y^l_{ij}\sim \varepsilon^{H(\hat\Phi_d)+H(\hat L_i)+H(\hat l^c_j)}$ and
$Y^d_{ij}\sim \varepsilon^{H(\hat\Phi_d)+H(\hat Q_i)+H(\hat d^c_j)}$
(rotation to the exact quark mass eigenstate basis does
not affect these order of magnitude estimates \cite{BGNN1}).
Most of the $L$-violating couplings in (\ref{Lviolmass})
are further suppressed with respect to the corresponding
Yukawa couplings. They can be  estimated as
\beq \label{lanum}
\lambda_{kij} \sim Y^l_{ij} \> \varepsilon^{H(L_k) - H(\Phi_d)}
            \sim \left(\frac{2 \sqrt{2} G_F}{\cos^2\beta}\right)^{1/2} \>
\> m_{l_i}\>{\varepsilon}^{H(l^c_j) - H(l^c_i) + H(L_k) -
H(\Phi_d)}\,,
\eeq
and
\beq  \label{lpnum}
\lambda^\prime_{kij} \sim Y^d_{ij} \varepsilon^{H(L_k) - H(\Phi_d)}
          \sim \left(\frac{2 \sqrt{2} G_F}{\cos^2\beta}\right)^{1/2} \>
 m_{d_i}\> {\varepsilon}^{H(d^c_j) - H(d^c_i) + H(L_k) - H(\Phi_d)}\,.
\eeq
These  equations show that
\vspace{-8pt}
\begin{itemize}\itemsep=-4pt
\item[($i$)]
the couplings $\la$ and $\lp$  involving fermions of the  third generation 

are respectively enhanced by $m_\tau$ and $m_b\,$;
\item[($ii$)]
like the lepton and down quark Yukawa couplings,
the $\lambda$, $\lambda^\prime$ couplings  increase with $\tan\beta\,$.
\end{itemize} \vspace{-8pt}

\noindent
In order to give a numerical estimate of the couplings,
we need a set of charges  and a value for the
$\cal H$-symmetry breaking parameter $\varepsilon\,$.
In the model discussed in  \cite{horizontal}, $\varepsilon\sim 0.22$ is
fixed by the magnitude of the Cabibbo angle, while the quark, lepton and
Higgs  charges are chosen to reproduce the values of the fermion masses
and CKM mixing angles. Besides reproducing the measured values,
the model  has some predictivity in the quark sector \cite{horizontal},
it yields estimates for ratios of neutrino masses \cite{YossiYuval,BGNN2},
and most important in the present context, it ensures that the
$L$-violating couplings in (\ref{Lviolmass}), 
(\ref{lanum}) and (\ref{lpnum})
are safely suppressed below the present experimental limits \cite{BGNN1}.
The following $\cal H$-charge assignments \cite{horizontal} fit the order
of magnitude of all the quark masses and CKM mixing angles
\beqa\label{Hquarks}
  \hat Q_{1}     \  \ \hat Q_{2}     \   \ \hat Q_{3}           &\qquad&
     \  \hat d^c_{1} \ \  \hat d^c_{2} \ \ \hat d^c_{3}   \qquad
     \  \hat u^c_{1} \ \ \hat u^c_{2} \ \ \hat u^c_{3}   \qquad
     \  \hat \Phi_{d}    \ \ \hat \Phi_{u}
\nnu\\
 (3) \ (2) \ (0) &\qquad&
 (3) \ (2) \ (2)  \qquad
 (3) \ (1) \ (0)   \qquad
 \ (0) \ (0) \,.
\eeqa
For the leptons, we will use two different sets of charges
that fit well the order of magnitude of the charged lepton masses.
We also use a different value for the squark masses
$ m_{\tilde q} \,$ for each set :
\beqa\label{Hleptons}
 \hat   L_{1}    \   \ \hat L_{2}     \   \ \hat L_{3}        &\qquad&
 \ \hat l^c_{1}\ \ \, \hat l^c_{2} \ \ \, \hat l^c_{3}   \qquad
\qquad m_{\tilde l}\>({\rm GeV}) \quad \ m_{\tilde q}\>({\rm GeV})\nnu\\
{\rm Model\ I: \ }\qquad\qquad  (4) \ (2) \ (0)    &\qquad&
  (4) \ (3) \ (3)    \qquad
\qquad \ 100   \qquad\qquad   170    \nnu\\
{\rm Model\ II:} \qquad\qquad  (3) \ (0) \ (0)    &\qquad&
  (5) \ (5) \ (3)    \qquad
\qquad \ 100   \qquad\qquad   350   \,.
\eeqa
The charge assignments and the sfermion mass values 
listed in (\ref{Hquarks}) and (\ref{Hleptons}), together with  
$\varepsilon\sim 0.22$ completely define the two models and allow
us to estimate the order of magnitude of the various decay rates. 
The charges of Model~I coincide with the charges of the
``master model'' of \cite{YossiYuval}. While in this model new physics
effects are induced dominantly by the new 
operators $\{O'\}$ in (\ref{setRp1})
arising from squark exchange, in Model~II the leading effects are due to
the scalar operators $\{O^S\}$ in (\ref{setRp2})
induced by slepton exchange.
The choices (\ref{Hquarks}) and (\ref{Hleptons}) for the horizontal charges
are not unique. Since the Yukawa interactions are invariant
under a set of  $U(1)$ symmetries such as hypercharge or lepton number,
 it is  always possible to shift the $H$-charges of any amount
proportional to one of the corresponding $U(1)$ quantum numbers,
without affecting the predictions for the masses and mixing
angles.\footnote{In models where the ${\cal H}$ symmetry is
gauged, shifts of this kind cannot be arbitrary but must respect
the constraints from anomaly cancellation.} In particular, a
shift proportional to $L\,$: $H(\hat L_i)\to H(\hat L_i) + n\,$,
$H( \hat l^c_i)
\to H( \hat l^c_i) - n\,$ and
$H(\hat\psi) \to H(\hat\psi)$ for all the other fields,
has the effect of suppressing (for $n>0\,$) all the $L$ violating couplings
in (\ref{Lviolmass}), (\ref{lanum}) and (\ref{lpnum})
by a factor of $\varepsilon^n\,$. 
We have found that already for $n=1$  the couplings are enough suppressed
so that no signal of new physics can be detected  in the
experimental quantities we are considering in this paper.
Notice also that Model~II can be derived from Model~I by means of  shifts
proportional to  lepton flavor numbers: $n_e = -1\,$, $n_\mu = -2\,$,
$n_\tau = 0\,$. This has the effect of enhancing some of the $\la$
couplings without affecting the charged lepton masses. Of course,
in Model~II the predictions for neutrino masses and mixings
will differ from the predictions of Model~I \cite{YossiYuval}.

\subsection{Coefficients and rates for the various decays }

In SUSY models without R-parity,
$b \to \tau$ decays can proceed through the exchange of
sleptons and/or squarks, yielding significant enhancements over
the SM rates.  In this section we give the expressions for the various
coefficients of the effective operators contributing to
the decays \Btn, \Btt,  \bqnn\ and \bstt\ including the new physics
contributions.

For the decay \Btn\ the coefficients appearing in (\ref{BlnAmpl})
and (\ref{BlnRate})  read
\beqa
{C_A^{\tau\bar\nu_i}}-{C_A^{\tau\bar\nu_i}}' & = & -
           \frac{{\lambda^{\prime}}^*_{31k} \lambda^{\prime}_{i3k}}
           {4\, G_{F}\, m^{2}_{\tilde d_k}} - \sqrt{2}\, V_{ub}
\nnu \\
C_P^{\tau\bar\nu_i} - {C_P^{\tau\bar\nu_i}}'& = & -
           \frac{{\lambda^\prime}^*_{k13}\lambda_{ki3}}
           { 2\, G_{F}\, m^{2}_{\tilde l_k}} \qquad\qquad (k\neq i)
\eeqa
where a sum over the repeated index $k$ is left understood.
The index $i$ refers to the final state neutrino
flavor which, as already said, can be different from $\nu_\tau$.

For the decay \Bqtt\ the coefficients in
(\ref{BqllAmpl}) and (\ref{BqllRate}) read
\beqa
 C^q_P  & = &  -\frac{\left(\lambda^*_{k33} \lambda^{\prime}_{k3q} +
                      \lambda_{k33} {\lambda^{\prime}}^*_{kq3}\right)}
                      {4\, G_F\, m^2_{\tilde l_k}}   \qquad (k\neq 3)
 \nnu \\
{C^q_P}'  & = &  \frac{\left(\lambda_{k33}{\lambda^{\prime}}^*_{kq3} -
                      \lambda^*_{k33}\lambda^{\prime}_{k3q}\right)}
                      {4\, G_F\, m^2_{\tilde l_k}}   \qquad (k\neq 3)
 \nnu \\
C^q_A & = & - \frac{{\lambda^{\prime}}^*_{3k3}\lambda^{\prime}_{3kq}}
             {8\, G_F\, m^2_{\tilde q_k}}\  +\ \Big[C^q_A\Big]_{\rm SM}
 \eeqa
where $\Big[C^q_A\Big]_{\rm SM}$ is given in  (\ref{CqASM}).

For the decay  $b\to  X_q\nu_{i}\bar \nu_{j}$ 
the coefficients appearing
in  (\ref{bqnnAmpl}) and  (\ref{bqnnRate}) read
\beqa
C^{qij}_L &=&  \frac{\lambda'_{i3k} {\lambda'}^*_{jqk}}
               {2\, G_F\, m^2_{\tilde d_k}} +
              \frac{1}{3}\, \Big[C^q_L\Big]_{\rm SM} \, \delta_{ij}
 \nnu \\
C^{qij}_R &=&  \frac{{\lambda'}^*_{ik3} \lambda'_{jkq}}
               {2\, G_F\, m^2_{\tilde d_k}}\,,
\eeqa
where $\Big[C^q_L\Big]_{\rm SM}$ is given in  (\ref{CqLSM}).
Finally,
for the decay \bstt, the coefficients of the new operators $\{O'\}$
($i=2,7,9,10$)
in (\ref{setRp1}) and $\{O^S\}$ in (\ref{setRp2}) at the scale
$ \tilde m = m_{\tilde l} $  are

\beqa  \label{RpCoef}
 C'_2\,(\tilde m) &=&  \frac{\lp^*_{k23}\lp_{k22}}
                       {2\, G_F\, m^2_{\tilde l_k} }
 \nnu \\
 C'_7 \, (\tilde m) &=& 0
 \nnu \\
 C^{\prime}_{9}\, (\tilde m)  \ = \
- C^{\prime}_{10}\, (\tilde m) & = & \frac{\lp^*_{3k3}\lp_{3k2}}{4\,
G_{F}\, m^2_{\tilde u_k} }
 \nnu \\
 C^S_9\, (\tilde m) \ = \ \  C^S_{10}\, (\tilde m) & = &
\frac{\lp_{k32}\la^*_{k33}}{2\, G_{F}\, m^2_{\tilde \nu_k}}
\qquad\qquad (k\neq 3)
 \nnu \\
{C^S_9}'\, (\tilde m) \ = \ - {C^S_{10}}'\, (\tilde m) & = &
\frac{\lp^*_{k23}\la_{k33}}{2\, G_{F}\, m^2_{\tilde \nu_k}}
\qquad\qquad (k\neq 3)\,.
 \eeqa
As regards the coefficients of the standard operators 
in (\ref{setSM}), they are affected by the new physics
only at the loop level. Since these are subleading effects we will 
neglect them. 

Few comments are in order. The antisymmetry in the
first two indices of the $\la$ couplings forces
$k\neq 3$ in the coefficients of the scalar operators.
In Model I, because of the
lepton charge assignments (\ref{Hleptons}) this
results in  a strong suppression of the scalar  couplings.
We have for example $C^S_9 \sim \varepsilon^4 \> C^{\prime}_{9}$
so that the leading new physics effects are due to the
operators induced by squark exchange.
In Model II, being $H(\hat L_2)=H(\hat L_3)=0$ there is no suppression
of the scalar couplings from the $\cal H$-symmetry. Then the
large value of the squark masses $m_{\tilde q} = 350\,$GeV
suppresses $C^{\prime}_{9}$ and $C^{\prime}_{10}$ down to
$\sim \varepsilon\> C^S_9$ so that in this model
slepton exchange gives the dominant effects.
As regards the \bsmm\ decay channel,
the charge difference $H(\hat L_2)-H(\hat L_3)=2$ of Model~I
implies a strong suppression of the $\lp\lp$ couplings.
In contrast, for the couplings involving the combination
$\lp\la$  the antisymmetry now allows $k=3\,$, with the noticeable
result that the scalar couplings in the $\mu$ channel are enhanced
with respect to the $\tau$ channel. Still the enhancement
is only of about a factor $ \varepsilon^{-2}$ so that 
we cannot expect particularly large new physics effects.
In Model~II there is no suppression of the squark exchange operators
for \bsmm\ with respect to the $\tau$ channel. Thus we can expect
that $C^{\prime}_{9}$ and $C^{\prime}_{10}$ will
still produce observable signals of new physics.
Finally, $C'_2$ that controls the relevance of the new long
distance contributions, is rather small  in both models. This is due to
the charge $H(\hat Q_2)=2$ which yields a relative
factor $\varepsilon^4 \,(m_{\tilde q}/m_{\tilde l})^2
\sim 0.001\,$(Model I)--$0.03\,$(Model II) with respect to the
$C^{\prime}_{9}$ short distance contribution. Therefore we can expect
that the resonance peaks will be smeared off by the dominant
new physics short distance effects.

%%%%%%%%%%%%%%%%%%%%%%%%%%%%%%%%%%%%%%%%

\section{Results }

The main results of our analysis consist of a set of numerical
predictions for several observables in the presence of new
physics from SUSY models without R-parity. We recall that
the theoretical framework we adopted is a straightforward
extension of successful models for fermion masses and CKM mixing
angles, and gives predictions for the R-parity breaking couplings
which are consistent with all the present experimental constraints
\cite{BGNN1}.

We have studied several observables measurable in
$b\to \tau\,,\nu_\tau$ and $b \to \mu$ transitions
within two different models, that were  defined in
(\ref{Hquarks}) and (\ref{Hleptons}).
In Model I, the horizontal charges
 coincide with  the charges of the ``master model''
discussed in  \cite{YossiYuval,horizontal}. The  sfermions masses
are $m_{\tilde l}=100\,$GeV and $m_{\tilde q}=170\,$GeV.
In this model the dominant new physics effects come from squark exchange
which generate the leading new effective operators.
Model~II  was introduced in order to study the effects of sleptons.
It is defined in terms  of a different set of lepton
horizontal charges  (\ref{Hleptons}). The sfermion masses
are $m_{\tilde l}=100\,$GeV and $m_{\tilde q}=350\,$GeV.

Varying the values of the new physics parameters
results in the following scaling behaviors of the branching ratios:
\vspace{-8pt}
\begin{itemize}\itemsep=-4pt
\item[($i$)]
In both models, the branching ratios scale as $(\tan\beta)^4\,$
and as $\varepsilon^{4\,n}\,$, where  $n$ is an arbitrary  shift
of the $H$-charges of the leptons proportional to lepton number
(see the discussion at the end of section IV-B).
Our results are given for $n=0$ and  $\tan\beta=1\,$.
\item[($ii$)]
In Model~I  the branching ratio for \bstt\ scales as
$(170\,{\rm GeV}/m_{\tilde q})^4$. Hence, for light squarks
($m_{\tilde q}\simeq 100\,$GeV) and moderate values of $\tan\beta$
($\gsim 2$) Model~I can predict values of the branching ratio up
to few $\times 10^{-5}\,$.
In Model~II  the rate for $\bstt$    scales as
$(100\,{\rm GeV}/m_{\tilde l})^4$ so that the main enhancement
with respect to our predictions (see table~I) can only come from
$\tan\beta > 1$.
\end{itemize} \vspace{-8pt}

Our results are collected in tables I and II, and in figures 1-8.

Table I lists the predictions for decays involving the transition
$b\to \tau\,,\, \nu_\tau$.  The first five lines give the results
for the decays \Btn, \Bqtt\ and \bxnn, while the
results for the branching ratio, forward-backward asymmetry and
longitudinal polarization asymmetries in \bstt\ are given in
the remaining entries. Table II collects the results for the corresponding
processes with final state  muons.
In both tables, the first column  lists the SM predictions
for the various observables (computed in the leading order approximation)
the second column lists the predictions of Model~I while
the results for Model~II are listed in the third column.

Since the two decays \bsmm\ and \bstt\ are affected by large
long distance effects, to single out  the short distance contributions
we have applied cuts on the dilepton invariant mass.
We study  \bstt\ in the region above the $\psi'$ ($\sh> 0.6$) while
\bsmm\ is analyzed below the resonance region ( $\sh < 0.4$).
A comparison  between the  total inclusive branching ratios
$\BR(\bsll)_{\scriptscriptstyle\rm no\> cut}\,$, the branching ratio
in the region within the cuts and the kinematic limits
$ \BR(\bsll)_{\scriptscriptstyle\sh<0.4\,(\sh>0.6) }$
and the short distance contribution in the same region
$ \BR^{\rm sd}(\bsll)$ shows the effects of the cuts on the total rates,
and their effectiveness in isolating the interesting contributions.

{}From the results in  table I, it is apparent that in most cases
the decays \Btn\ and \bqnn\ are not very sensitive to the
sources of new physics we are analyzing here. In  both our
models these decays do not show any significant
enhancement with respect to the SM rates. New physics from
Model~I can enhance the rates for \Bstt\ and \Bdtt.
However, the rates remain very small, and because of
the large theoretical uncertainty related to $f_B\,$ it is
not obvious that these signals could be unambiguously identified.  
In Model~II the
branching ratio for \Bstt\ increases by two orders of magnitudes,
up to $\sim 10^{-4}$. This process cannot be searched for at
$B$-factories running at the $\Upsilon(4S)\,$. However, hadron
colliders might be able to detect this enhancement, depending on the
efficiency in identifying the $\tau$'s. In this model we also observe a
two orders of magnitude  enhancement for \Bdtt\ which appears more
promising for new physics searches at future $B$-factories.

In the muon channel, the corresponding decays \Bqmm\ are
also sensitive to new physics effects from both models.
However, even if the decay rates are
enhanced by two orders of magnitude, the branching ratios are still
only at the level of $\sim 7 \times 10^{-7}$ for $B_s$ and
$\sim 3\times 10^{-8}$ for $B_d$.

As regards the decay \bsmm, we see that no signal of new physics
is expected in Model I.  The rate, the forward-backward
asymmetry and the longitudinal polarization  asymmetry
remain at their SM values. This decay is somewhat more sensitive to
new physics from Model II, which induces a factor of two enhancement
of the short distance contributions. However, as it clear also by
inspecting figs.~6~-~8, below the cut $\sh < 0.4$ 
the overall effects are very 
likely too small to be unambiguously identified above the theoretical and
experimental uncertainties.

We turn now to the decay \bstt. In both our models the branching
ratio for this decay is enhanced by more than one order of
magnitude, at a level that could be observable with a good $\tau$
identification efficiency. Notice that with our choice 
of new physics parameters,
Model~I and Model~II both
predict very similar rates (see also fig.~1) even if the
respective enhancements  are induced by effective operators of
quite a different nature. 
Therefore, even if such a large signal of new
physics will be observed, it would not be possible to identify which
kind  of new physics is producing the effect just from a
measurement of the decay rate. In contrast, we see that 
the forward-backward asymmetry  $A_{\rm FB}$ (fig.~2)
and the longitudinal polarization asymmetry 
$\lop$ (fig.~3) could provide the additional information needed to 
disentangle the different effects.

In Model~I $A_{\rm FB}$ is almost at the  20\% level, and opposite in 
sign with respect to the SM in the whole kinematic region. 
In contrast, in Model~II this asymmetry is vanishingly small. 
On the other hand in Model~II the longitudinal 
component $\lop$ is rather large, about twice 
the SM prediction, while in Model~I  it remains 
close to the SM value.

As regards the other two polarization components, 
in both models the transverse asymmetry $\trp$ is about a factor of three 
smaller than in the SM.  The T-odd 
component $\nop$ which in the SM is at the 5\% level, 
is practically zero in both our models. This can be traced back  to the 
fact that the new (real) short distance contributions dominate 
over the absorptive part of the decay rate related to 
on shell $\bar c c$ intermediate states. 

\section{Conclusions}

In this paper we have studied rare $b$ decays into
leptons of the third generation. These decay modes
are well suited to study sources of new physics
which couple more strongly to third family. 
We have first discussed a general
framework for studying effects beyond the SM, and we 
have introduced general four-fermion amplitudes
for the decays \Btn, \Bqtt\ and \bqnn. We have also 
defined the effective Hamiltonian for the decay
\bstt\ in terms of an enlarged operator basis.

We have applied our results to the study
of SUSY models without R-parity and without $L$ number.
In these models new contributions to the decays appear already
at the tree level, through  new effective operators
generated by squark and slepton exchange having
a different structure than the SM ones. In order
to derive numerical predictions for the various observables,
we have embedded SUSY without R-parity in the framework of
models for fermion masses based on Abelian horizontal symmetries.
This allowed us to estimate the order of magnitude of the
various R-parity violating couplings.

We have carried out a numerical study of two representative models,
in which new physics effects arise from two different sets of effective
operators, induced respectively by squarks and by sleptons exchange.
We found that  the most sensitive among the processes involving the
$b\to \tau$ transition  are the decays \Bqtt\ ($q=d,s$) and \bstt, 
that can be enhanced up  to two orders of magnitude over the SM rates. 
If such an enhancement is observed,
additional measurements of the forward-backward asymmetry and of 
the $\tau$ longitudinal polarization in \bstt\ 
can be very helpful in identifying the kind of underlying new physics.
We have confronted the predictions for decays into final state 
taus with the corresponding results for decays into muons.
The decay \bsmm\ is not sensitive to this kind of new physics,  
and we found that only  the  \Bqmm\ decay modes show  
enhancements comparable to the $\tau$ channel. 
However, even with the new physics contributions, 
the overall rates for  these decays remain rather small.

\vspace{1.5cm}

\acknowledgements
We thank Yuval Grossman, Zoltan Ligeti and Yossi Nir for
useful comments on the manuscript.
We also acknowledge the Particle Theory Group of the Weizmann
Institute of Science, where this research was carried out,
for the hospitality and for the pleasant working atmosphere.

\vspace{1.5cm}

% \appendix
\appendix\section{Input parameters}

\label{input}
\parbox{13cm}{
\beqa
&& m_b=4.8\ \GeV,\ m_c= 1.4\ \GeV,\ m_s=0.2\ \GeV,\ m_t=176\ \GeV\ , \nnu\\
&& m_{\m} = 0.106\ \GeV,\ m_{\t} = 1.777\ \GeV,\ M_W=80.2\ \GeV,\  \nnu\\
&& V_{tb}^{}=1,\ V_{ts}=V_{cb}=-0.040,\ V_{td}^{}=0.009,\
   V_{ub}=0.003,\ \nnu\\
&&  f_{B_d} = 200\, {\rm MeV}, \ f_{B_s} = 230\, {\rm MeV},\
    m_B = 5.3\, {\rm GeV},\ \tau_B=1.6\, {\rm ps}\, \nnu \\
&& \Lambda_{\rm{QCD}} = 225\ \MeV,\ \mu=m_b,\ \alpha(M_Z)=1/129,\
\sin^2{\theta_{\rm{W}}} = 0.23,\ \nnu\\
&&  \BR(B\to X_c\,l\bar{\n}_l)= 10.4\%\, . \nnu
\eeqa }

 }  %%% end tighten  %%  For 35 or 27 pages preprints.

\newpage

{\tighten

}  % end tighten

%%%%%%%%%%%%%%%%%%%%%%%%%%%%%%%%%%%%%%%%%%%%%%%%%%%%%%%%

\newpage
%%%%%%%%%%% TABLE 1 %%%%%%%%%%%%%%%%%%%%%%%%%%%%%%
%
\begin{table}[t]
\begin{center}
 {\tighten
\caption{\baselineskip 16pt
Predictions for the various decay rates and asymmetries for
$b \to \tau$ transitions in the standard model and in the R-parity
violating models  discussed in the text.
Model~I is sensitive to  operators generated by squark exchange. The
lepton horizontal charges are $ H(\hat L) = (4 \,, 2 \,, 0) \,$,
$ H(\hat   l^c) = (4 \,, 3 \,, 3) $  while the
SUSY masses are $m_{\tilde l} =100$ GeV and 
$m_{\tilde q} =170$ GeV. Model~II is sensitive to  operators generated by
slepton exchange, with  horizontal charges
$ H(\hat L)  = (3 \,, 0 \,, 0) \,$, $ H(\hat l^c ) = (5 \,, 5 \,, 3)$
and SUSY masses $m_{\tilde l} =100$ GeV and  $m_{\tilde q} =350$ GeV.
In both models the value of the horizontal symmetry breaking parameter 
is $\varepsilon = 0.22\,$. }
\label{table1}
\vspace{0.3cm}
\begin{tabular} {c  c  c  c }
Process & Standard Model  &  Model 1   &  Model 2 \\
  \hline
\noalign {\vspace{1truemm} }
$ \BR\ (B^-\to \, \tau^-\> \bar\nu\>)
             $&$  7.1 \times 10^{-5} $&$ 7.2\times 10^{-5} $
& $  7.4\times 10^{-5} $ \\
$ \BR\ (B_s\to \, \tau^+\tau^-)
             $&$    9.1 \times 10^{-7} $&$ 5.7\times 10^{-6}$
& $  1.8\times 10^{-4} $ \\
$ \BR\ (B_d\to \, \tau^+\tau^-)
             $&$    4.3 \times 10^{-8} $&$ 1.9\times 10^{-7}$
& $  6.3\times 10^{-6} $ \\
$  \BR\ (b\to X_s\> \nu\>\bar{\nu}\>)
             $&$  4.4 \times 10^{-5} $&$ 6.7\times 10^{-5}$
& $  5.0\times 10^{-5} $ \\
$  \BR\ (b\to X_d\> \nu\>\bar{\nu}\>)
             $&$  2.7 \times 10^{-6} $&$ 3.9\times 10^{-6}$
& $  3.0\times 10^{-6}   $ \\
\noalign {\vspace{1truemm} }
\hline
\noalign {\vspace{1truemm} }
$ \BR\ (b\to X_s\, \tau^+\tau^-)_{\scriptscriptstyle\rm no\> cut}
           $&$ 4.9 \times 10^{-6} $&$ 9.6 \times 10^{-6}$
& $  1.0 \times 10^{-5} $\\
$ \BR\ (b\to X_s\, \tau^+\tau^-)_{\scriptscriptstyle\sh>0.6}
           $&$ 1.5 \times 10^{-7} $&$ 4.1\times 10^{-6}$
& $   4.6 \times 10^{-6} $\\
$ \BR^{\rm sd}(b\to X_s\, \tau^+\tau^-)_{\scriptscriptstyle\sh>0.6}
           $&$ 1.6 \times 10^{-7} $&$ 4.1 \times 10^{-6}$
& $   4.6 \times 10^{-6} $\\
$ \left\langle A_{\rm FB}^\tau \right\rangle_{\scriptscriptstyle\sh>0.6}
           $&$\> -0.13 $&$ \  0.18 $&$ -0.03 $\\
$ \left\langle\>{\cal P}_{\rm L}^\tau\>
\right\rangle_{\scriptscriptstyle\sh>0.6}
           $&$   -0.34 $&$   -0.40 $&$ -0.68 $\\
$ \left\langle\>{\cal P}_{\rm T}^\tau\>
\right\rangle_{\scriptscriptstyle\sh>0.6}
           $&$   -0.40 $&$   -0.13 $&$ -0.14 $\\
$ \left\langle\>{\cal P}_{\rm N}^\tau\>\right
\rangle_{\scriptscriptstyle\sh>0.6}
           $&$\>\ 0.05 $&$\>\ 0.00 $&$\ 0.01 $\\
\noalign {\vspace{1truemm} }
\end{tabular}
 }     % end tighten
\end{center}
\end{table}
%

%%%%%%%%%%% TABLE 2 %%%%%%%%%%%%%%%%%%%%%%%%%%%%%%
%
\begin{table}[b]
\begin{center}
 {\tighten
\caption{\baselineskip 16pt
Predictions for the various decay rates and asymmetries for
$b \to \mu$ transitions in the standard model and in the R-parity violating
models discussed in the text.
Model~I is sensitive to  operators generated by squark exchange. The
lepton horizontal charges are $ H(\hat L) = (4 \,, 2 \,, 0) \,$,
$ H(\hat   l^c) = (4 \,, 3 \,, 3) $  while the
SUSY masses are $m_{\tilde l} =100$ GeV and 
$m_{\tilde q} =170$ GeV. Model~II is sensitive to  operators generated by
slepton exchange, with  horizontal charges
$ H(\hat L)  = (3 \,, 0 \,, 0) \,$, $ H(\hat l^c ) = (5 \,, 5 \,, 3)$
and SUSY masses $m_{\tilde l} =100$ GeV and $m_{\tilde q} =350$ GeV. 
In both models the value of the horizontal symmetry breaking parameter 
is $\varepsilon = 0.22\,$. }
\label{table2}
\vspace{0.3cm}
\begin{tabular} {c  c  c  c }
Process & Standard Model  &  Model 1   &  Model 2 \\
  \hline
\noalign {\vspace{1truemm} }
$ \BR\ (B^-\to \, \mu^-\> \bar\nu\>)
         $&$  3.2 \times 10^{-7} $&$ 3.2\times 10^{-7} $
& $  3.3\times 10^{-7} $ \\
$ \BR\ (B_s\to \, \mu^+\mu^-)
         $&$    4.3 \times 10^{-9} $&$ 7.9\times 10^{-7}$
& $  7.2\times 10^{-7} $ \\
$ \BR\ (B_d\to \, \mu^+\mu^-)  $&$  2.1 \times 10^{-10}
         $&$ 2.9\times 10^{-8} $&$  2.7\times 10^{-8} $ \\
\noalign {\vspace{1truemm} }
\hline
\noalign {\vspace{1truemm} }
$ \BR\ (b\to X_s\, \mu^+\mu^-)_{\scriptscriptstyle\rm no\> cut}
             $&$ 3.1 \times 10^{-4} $&$ 3.1\times 10^{-4}$
& $  3.4 \times 10^{-4} $  \\
$ \BR\ (b\to X_s\, \mu^+\mu^-)_{\scriptscriptstyle\sh<0.4 }
             $&$ 4.3 \times 10^{-6} $&$ 4.5 \times 10^{-6}$
& $   8.3 \times 10^{-6} $\\
$ \BR^{\rm sd}(b\to X_s\, \mu^+\mu^-)_{\scriptscriptstyle\sh<0.4}
             $&$ 3.9 \times 10^{-6} $&$ 4.1 \times 10^{-6}$
& $  7.7 \times 10^{-6} $\\
$ \left\langle A^\mu_{\rm FB}   \right\rangle_{\scriptscriptstyle\sh<0.4}
             $&$\> -0.01 $&$\>\ 0.00 $&$\ 0.08 $\\
$ \left\langle\> {\cal P}_{\rm L}^\mu\>
\right\rangle_{\scriptscriptstyle\sh<0.4}
             $&$ -0.57 $&$ -0.56 $&$ -0.73 $ \\
\noalign {\vspace{1truemm} }
\end{tabular}
 }     % end tighten
\end{center}
\end{table}
%

%%%%%%%%%%%%%%%%%%%%%%%%%%%%%%%%%%%%%%%%%%%%%%%%%%%%%%%%

\newpage

\begin{figure}[tt]
\epsfxsize=10truecm
\centerline{\epsffile[100 203 503 503]{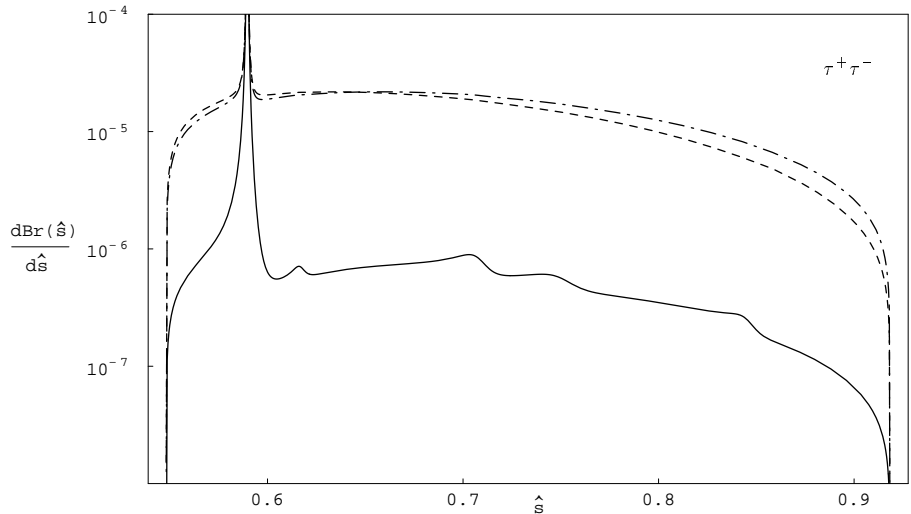}}
\caption{Predictions for the differential 
branching ratio $ \BR(\bstt) $  as a
function of $ \hat{s} $ in the standard model (solid), 
in Model~I (dashed)
and in Model~II (dash-dotted) discussed in the text, 
including the long distance  contribution.
Model~I is sensitive to  operators generated by squark exchange. The
lepton horizontal charges are $ H(\hat L) = (4 \,, 2 \,, 0) \,$,
$ H(\hat   l^c) = (4 \,, 3 \,, 3) $  while the
SUSY masses are $m_{\tilde l} =100$ GeV and 
$m_{\tilde q} =170$ GeV. Model~II is sensitive to  operators generated by
slepton exchange, with  horizontal charges
$ H(\hat L)  = (3 \,, 0 \,, 0) \,$, $ H(\hat l^c ) = (5 \,, 5 \,, 3)$
and SUSY masses $m_{\tilde l} =100$ GeV and $m_{\tilde q} =350$ GeV. 
In both models the value of the horizontal symmetry breaking parameter 
is $\varepsilon = 0.22\,$.
  }\end{figure}

\begin{figure}[b]
\epsfxsize=10truecm
\centerline{\epsffile[100 203 503 503]{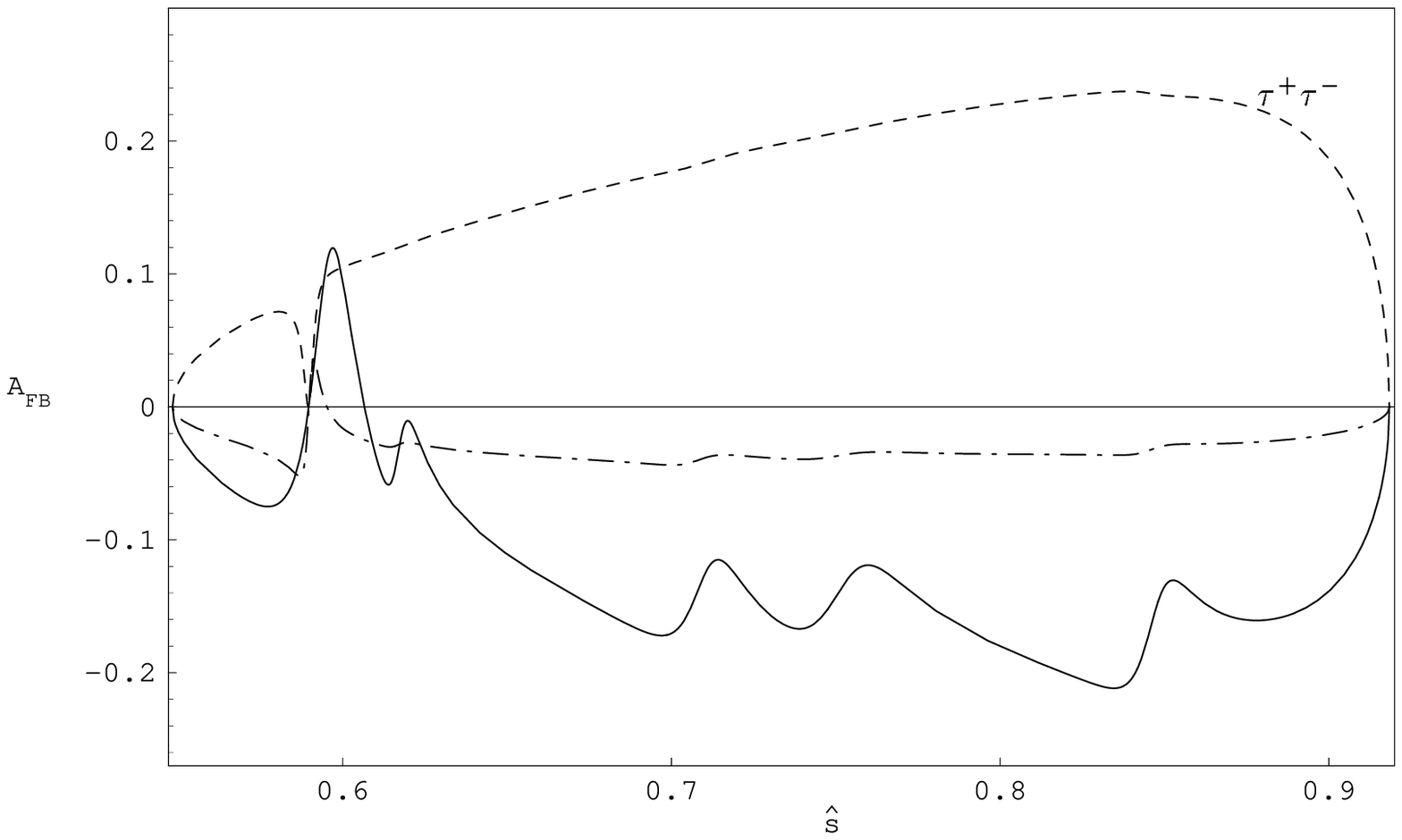}}
\caption{Predictions for the forward-backward asymmetry $ A_{\rm FB} $  
for the $\tau$ lepton  as a function of $ \hat{s} $ 
in the standard model (solid), in Model~I (dashed)
and in Model~II (dash-dotted) discussed in the text.
The new physics model parameters are as  in  fig.~1. 
  }\end{figure}

%%%%%%%%%%%%%%%%%%%%%%%%%%%%%%%%%%%%%%%%%%%%%%%%%%%%%%%%

\newpage

\begin{figure}[t]
\epsfxsize=10truecm
\centerline{\epsffile[100 203 503 503]{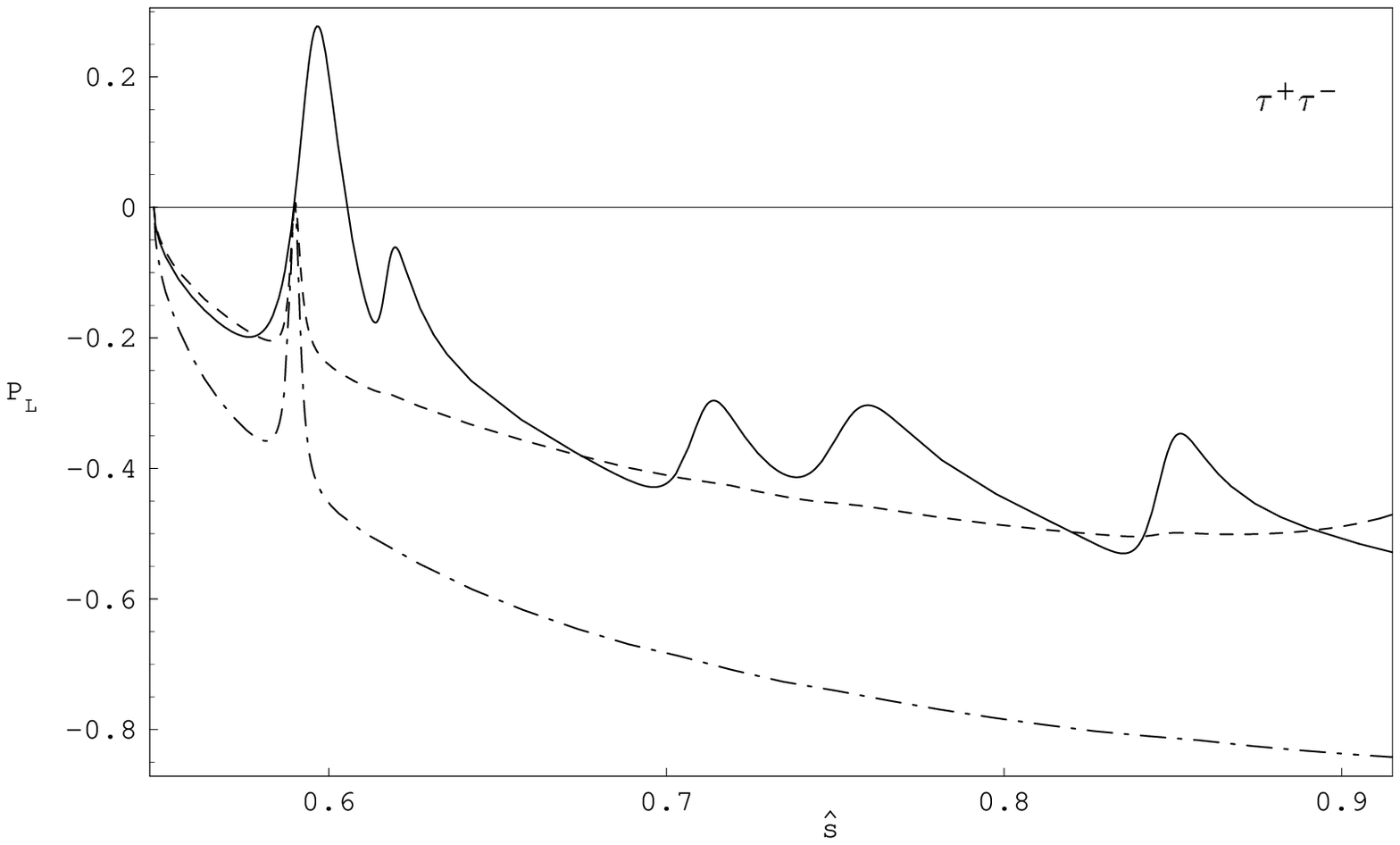}}
\caption{Predictions for the longitudinal polarization $ {\cal P}_{\rm L} $  
for the $\tau$ lepton  as a function of $ \hat{s} $ in the standard 
model (solid),
in Model~I (dashed) and in Model~II (dash-dotted)
discussed in the text.
The new physics model parameters are as in fig.~1. 
  }\end{figure}

\begin{figure}[b]
\epsfxsize=10truecm
\centerline{\epsffile[100 203 503 503]{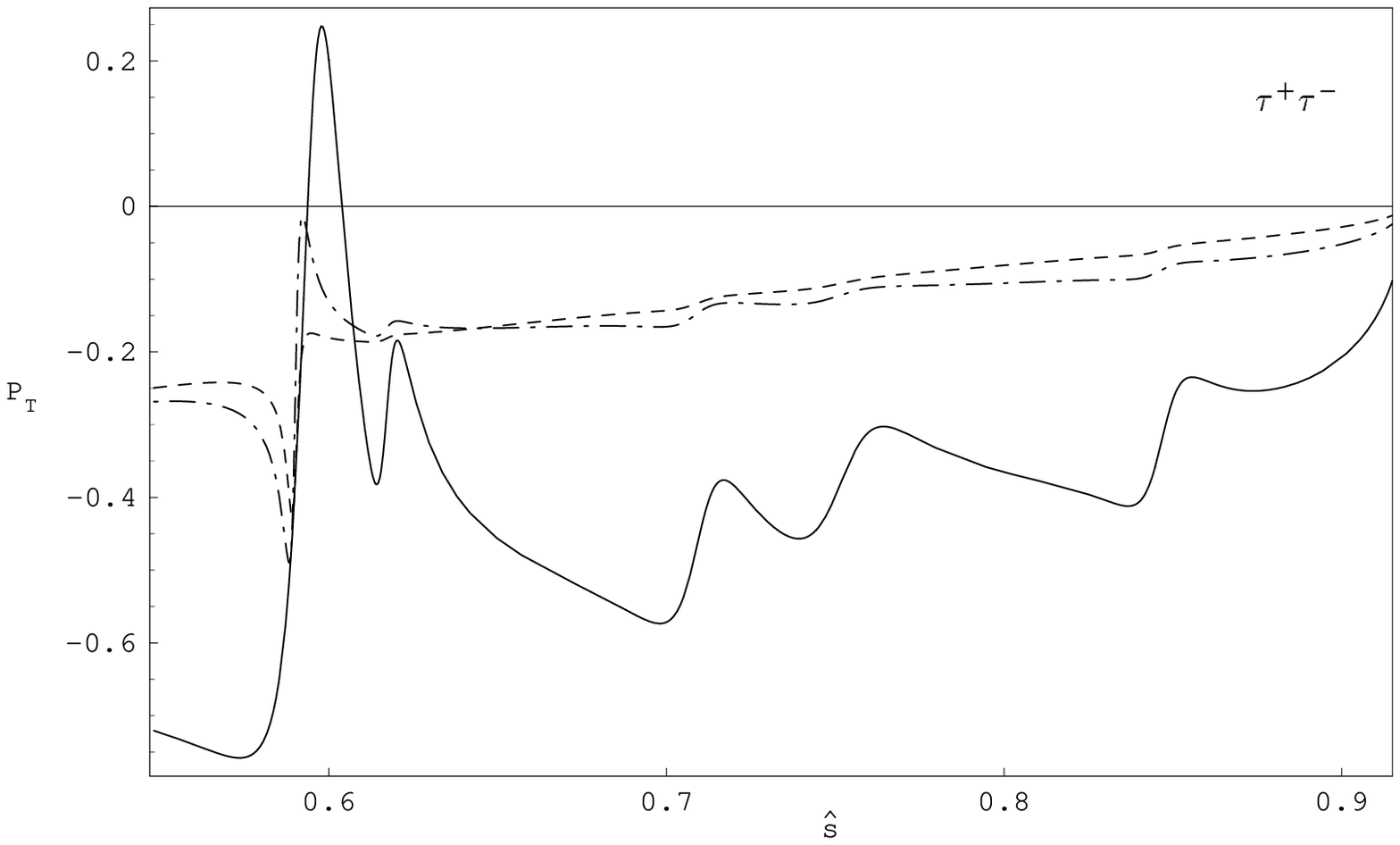}}
\caption{Predictions for the transverse polarization $ {\cal P}_{\rm T} $  
for the $\tau$ lepton  as a function of $ \hat{s} $ in the 
standard model (solid),
in Model~I (dashed) and in Model~II (dash-dotted) discussed in the text.
The new physics model parameters are as  in  fig.~1. 
  }\end{figure}

%%%%%%%%%%%%%%%%%%%%%%%%%%%%%%%%%%%%%%%%%%%%%%%%%%%%%%%%

\newpage

\begin{figure}[t]
\epsfxsize=10truecm
\centerline{\epsffile[100 203 503 503]{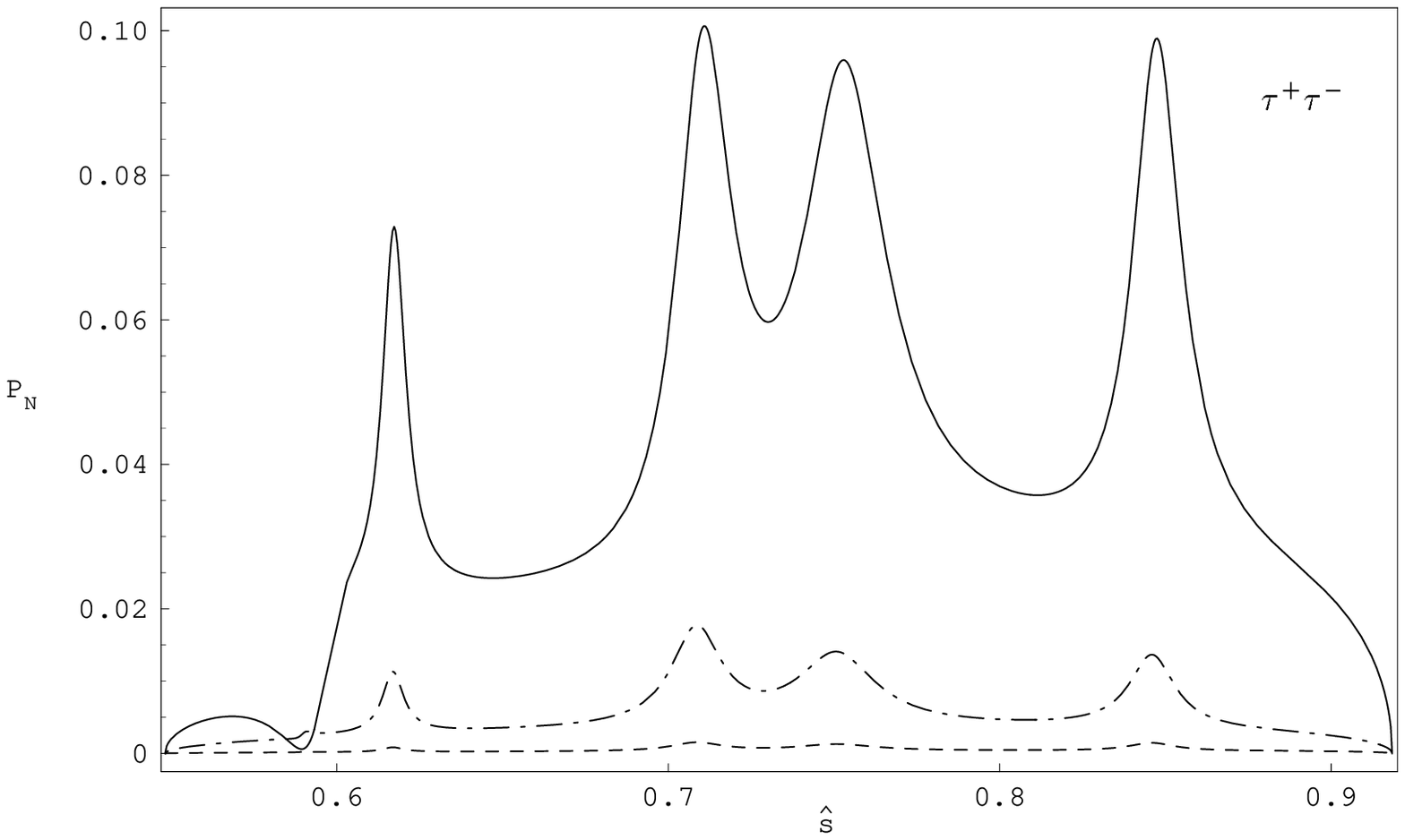}}
\caption{Predictions for the normal polarization 
$ {\cal P}_{\rm N} $  for the $\tau$ lepton
as a function of $ \hat{s} $ in the 
standard model (solid), in Model~I (dashed)
and in Model~II (dash-dotted) discussed in the text.
The new physics model parameters are as  in  fig.~1. 
  }\end{figure}

\begin{figure}[b]
\epsfxsize=10truecm
\centerline{\epsffile[100 203 503 503]{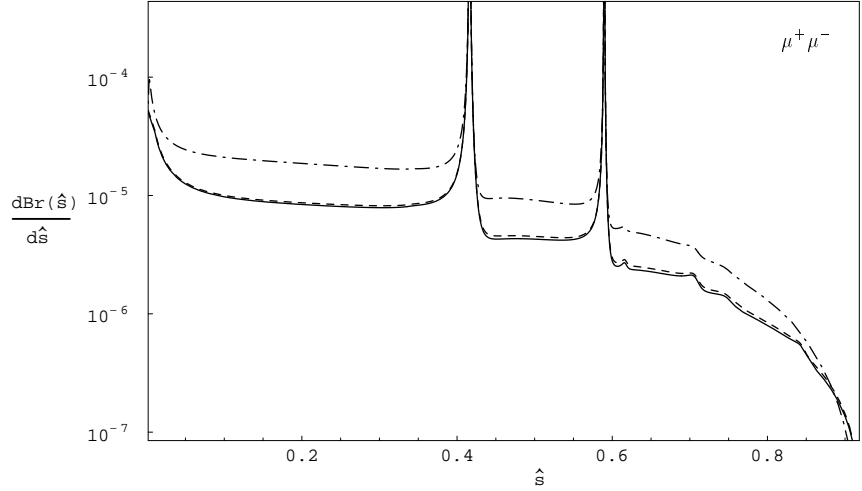}}
\caption{Predictions for the differential branching ratio $ \BR(\bsmm) $  
as a function of $ \hat{s} $ in the standard model (solid), in Model~I 
(dashed) and in Model~II (dash-dotted) discussed in the text, 
including the long distance  contribution.
The new physics model parameters are as in  fig.~1. 
  }\end{figure}

%%%%%%%%%%%%%%%%%%%%%%%%%%%%%%%%%%%%%%%%%%%%%%%%%%%%%%%%

\newpage

\begin{figure}[t]
\epsfxsize=10truecm
\centerline{\epsffile[100 203 503 503]{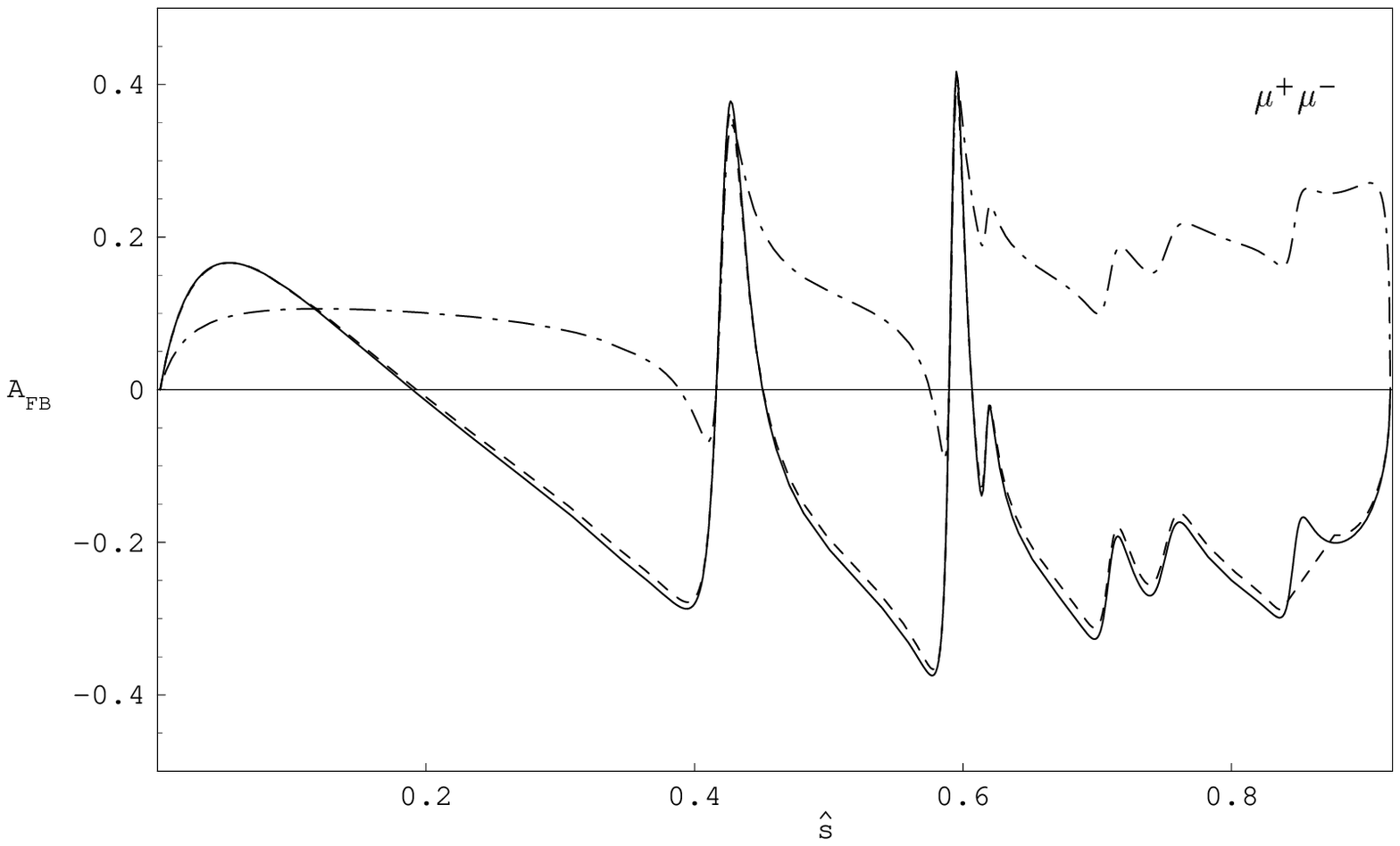}}
\caption{Predictions for the forward backward asymmetry 
$ A_{\rm FB} $  for the $\mu$
lepton  as a function of $ \hat{s} $ in the standard model 
(solid), in 
Model~I (dashed) and in Model~II (dash-dotted) discussed in the text.
The new physics model parameters are as in  fig.~1. 
  }\end{figure}

\begin{figure}[b]
\epsfxsize=10truecm
\centerline{\epsffile[100 203 503 503]{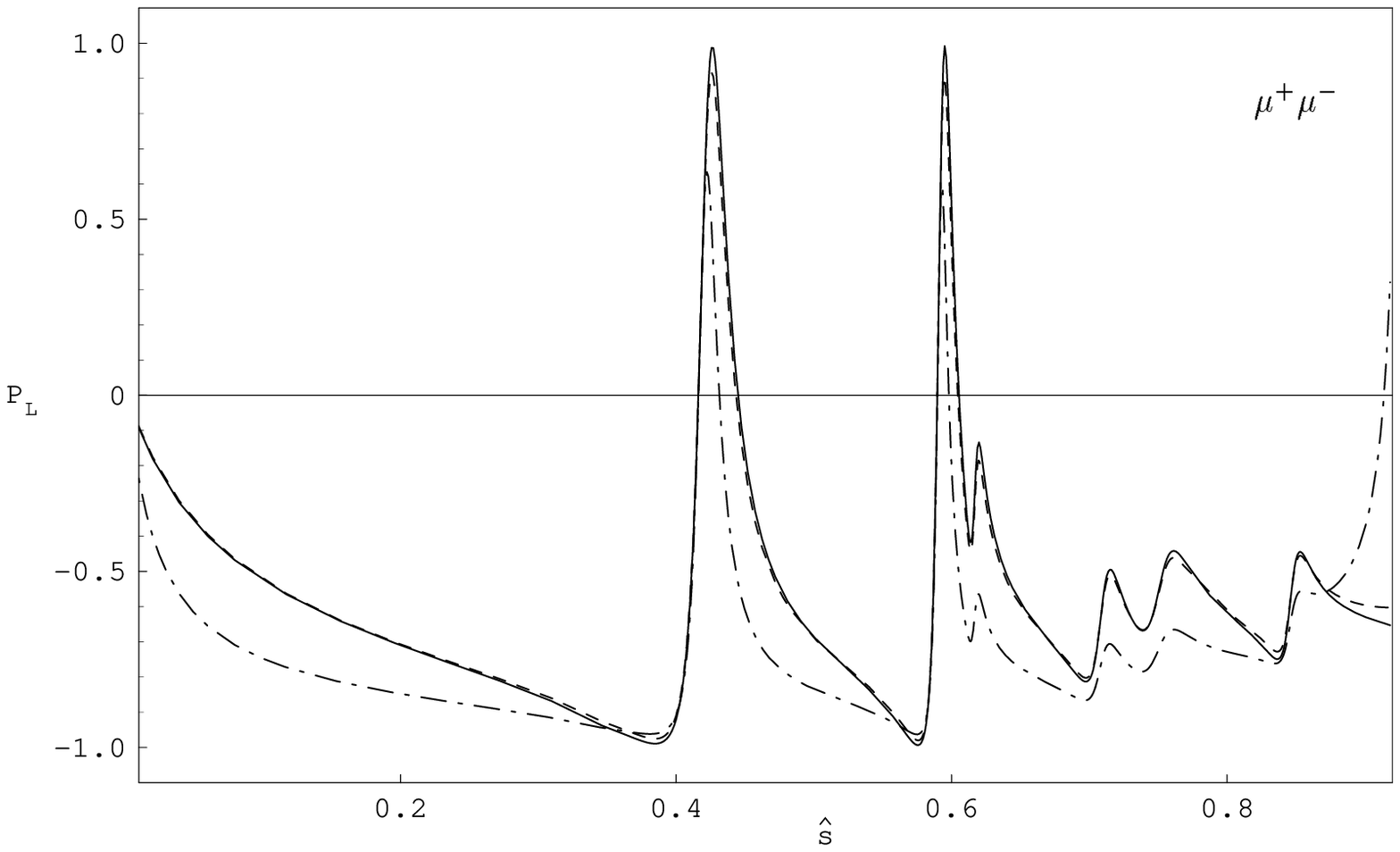}}
\caption{Predictions for the longitudinal polarization $ {\cal P}_{\rm L} $
for the $\mu$ lepton  as function of $ \hat{s} $ in the 
standard model (solid),
in  Model~I (dashed) and in Model~II (dash-dotted) discussed in the text.
The new physics model parameters are as in  fig.~1. 
  }\end{figure}

\end{document}